\documentstyle[tighten,eqsecnum,aps,psfig]{revtex}

\begin{document}
\draft

\newcommand{\q}{$q$}
\newcommand{\x}{$x$}
\newcommand{\s}{$S$}
\newcommand{\pot}{$\varphi$}
\newcommand{\mq}{q}
\newcommand{\mx}{x}
\newcommand{\ms}{S}
\newcommand{\be}{\begin{equation}}
\newcommand{\ee}{\end{equation}}
\newcommand{\bea}{\begin{eqnarray}}
\newcommand{\eea}{\end{eqnarray}}

\title{The Cosmological Mass Function with 1D Gravity}

\author{Pierluigi Monaco} 
\address{Institute of Astronomy, University of Cambridge, Madingley
Road, Cambridge CB3 0HA -- GB\\ Dipartimento di Astronomia,
Universit\`a di Trieste, via Tiepolo 11, 34131 Trieste -- Italy\\
email monaco@ast.cam.ac.uk}

\author{Giuseppe Murante} 
\address{Dipartimento di Fisica, Universit\`a di Milano, Via Celoria
16, 20133 Milano -- Italy\\ email Giuseppe.Murante@uni.mi.astro.it}

\date{\today}
\maketitle
\begin{abstract}
The cosmological mass function problem is analyzed in full detail in
the case of 1D gravity, with analytical, semi-analytical and numerical
techniques.  The extended Press \& Schechter\cite{ps74} theory is
improved by detailing the relation between smoothing radius and mass
of the objects.  This is done by introducing in the formalism the
concept of a growth curve for the objects.  The predictions of the
extended Press \& Schechter theory are compared to large N-body
simulations of flat expanding 1D universes with scale-free power
spectra of primordial perturbations.  The collapsed objects in the
simulations are located with a clump-finding algorithm designed to
find regions that have undergone orbit crossing or that are in the
multi-stream regime (these are different as an effect of the finite
size of the multi-stream regions).  It is found that the
semi-analytical mass function theory, which has no free parameters, is
able to recover the properties of collapsed objects both statistically
and object by object.  In particular, the predictions of regions in
orbit crossing are optimized by the use of Gaussian filtering, while
the use of sharp $k$-space filtering apparently allows to reproduce
the larger multi-stream regions.  The mass function theory does not
reproduce well the clumps found with the standard friends-of-friends
algorithm; however, the performance of this algorithm has not been
thoroughly tested in the 1D cosmology.  Our preliminary analyses of
the 3D case confirms that the techniques developed in this paper are
precious in understanding the cosmological mass function problem in
3D.
\end{abstract}

\widetext

\section{Introduction}

The most widely accepted view of the formation and evolution of the
Large Scale Structure of the Universe (LSS) is currently based upon
the gravitational instability of a self-gravitating density field.
Such a density field should be dominated by non baryonic,
collisionless dark matter\cite{cl96}.  The density field is supposed
to be homogeneous, apart from small random Gaussian fluctuations which
trigger the instability. The evolution of these fluctuations under the
action of self-gravity is described by the Vlasov-Poisson
equations\cite{p80}. Analytical approaches, typically linear or
(Lagrangian or Eulerian) perturbative approximations, are able to
reproduce the evolution of the matter field to the fully-developed
non-linear stage only in case of special symmetries.  Otherwise the
validity of perturbative approaches is limited to the linear or
so-called mildly non-linear regimes, when the evolution is still
laminar and no {\it orbit crossing}\footnote{Orbit crossing is said to
take place when two or more infinitesimal matter elements flow from
different Lagrangian position to the same Eulerian coordinate.}  has
occurred.

The evolution of the density fluctuations in the highly non-linear
regime is usually described by means of large numerical simulations,
in which the density field is sampled by massive particles
accelerating according to their Newtonian gravitational force.  These
numerical N-body simulations, which are affected by severe numerical
resolution limits, are able to provide final configurations which are
not trivial to analyze and understand in detail.  A joint use of
numerical simulations and analytical approximations seems the best way
to deepen our understanding of the gravitational problem in cosmology.

It is well known that a density field with random (Gaussian
distributed) small perturbations gives birth to a hierarchy of
collapsed, relaxed structures.  One important quantity which
characterizes the properties of these collapsed structures is the
distribution of their masses, usually called the mass function
(hereafter MF).  An accurate MF theory is the starting point for
modeling the statistical behavior of most cosmological objects, and
can be extended to give predictions of the main properties of
dark-matter structures.

The MF problem has been recently reviewed by Monaco (1998)\cite{m98}.
While most predictions, since the one of Press \& Schechter
(1974\cite{ps74}, hereafter PS; but see also Doroshkevich
1967\cite{d67}), were based on heuristic extrapolations of linear
theory, it is possible to construct a MF theory based on the powerful
Lagrangian perturbation theory\cite{lpt,b94,bch95,c95}; this is shown
in Monaco (1997a,b)\cite{m97a,m97b}.  A key point is that the same
basic concepts of collapse and of total mass of an object are not well
defined.  Collapse may be confused with total or partial
virialization, and the total mass of the objects is usually estimated
through order-of-magnitude arguments.  Within the Lagrangian
perturbation framework, it is possible to define collapse in a precise
way, i.e. as orbit crossing (hereafter OC) of mass elements.  In 3D,
this definition does not distinguish between just-collapsed
structures, like two-dimensional or one-dimensional transients
(`pancakes' or `filaments'), and fully virialized clumps.  It is then
appropriate to compare theory and numerical simulations in terms of
regions which have undergone OC, then checking the relation between
these and the collapsed clumps found with other, more standard
algorithms.  Besides, the problem of a clear definition of the total
mass of a structure is of a geometrical nature.  It is necessary to
determine the volume and topology of those extended regions that share
the property of `being collapsed'.  We will propose in this paper a
simple and effective way to address this problem.

As an intermediate step to the full solution of the MF problem, it is
convenient to analyze a simplified case, that of 1D gravity.  While
being a purely academic problem, 1D gravity provides a valuable
gymnasium in which to develop and test analytical predictions.  The 1D
problem was already analyzed, in the framework of the adhesion
theory\cite{gss89}, by Doroshkevich \& Kotok (1990)\cite{dk90} and by
Williams et al. (1991)\cite{whp91}, while Doroshkevich et
al. (1980)\cite{detal80} studied the problem with N-body simulations.
From the dynamical point of view, the geometry of first collapse
(which coincides with OC) is predicted to be 1D, or
pancake-like\cite{z70,sz89} (see in particular Shandarin et al.
1995\cite{smm95}), so 1D is the relevant restricted case, especially
when OC is used as collapse definition.  As will be discussed in
Section II.A, Lagrangian perturbation theory is trivial in 1D because
the first-order Zel'dovich approximation is exact up to OC\cite{sz89}.
By comparing theory with simulations, it is then possible to test the
elements which are missed by this approach, without taking into
account the effects of the truncation of the perturbative series.  The
geometry of the 1D structures is highly simplified because connected
regions are just segments.  The statistics of collapsed regions are
also greatly simplified.  It will be shown in Section II.A that the
statistics of collapsed regions can be based on a Gaussian field,
although the gravitational evolution of perturbations induces strong
non-Gaussianities in the density distribution.

Anyway, the MF problem in 1D is not trivial and presents many
complications relevant to the 3D problem.  The reason why the MF
problem can benefit from a 1D analysis is that it is, in some sense,
1D by itself.  It was shown by Monaco (1997b)\cite{m97b} that, in the
general case in which the non-linear and non-local gravitational
dynamics are taken into account, the MF problem can be formulated in
terms of an infinite-dimensional diffusion system (see Section II.B
for more details) that can be `projected' onto one dimension and
solved as a random walk with white or colored noise (Peacock \&
Heavens 1990, hereafter PH\cite{ph90}; Bond et al. 1991\cite{bce91}).
Also the merging histories of dark-matter halos are commonly found by
solving a 1D diffusion equation (Bond et al. 1991\cite{bce91}; Lacey
\& Cole 1993\cite{lc93}).  This reflects the fact that only a limited
amount of information is present in the MF, to the extent that limited
information on a structure is given by its mass.  The clustering of
collapsed halos, for instance, requires knowledge of the space
correlations\cite{pml98}, and thus would not benefit much from a 1D
analysis.  The utility of the results of this paper for a 3D analysis
of the mass function problem will be discussed in the Conclusions.

The scope of this paper is to give a complete approximated solution of
the MF problem in 1D, and to compare it to the results of numerical
simulations.  The concepts and techniques developed here can be
adapted to the full 3D case; this will be discussed at the end of
Section V.  The plan of the paper is the following. In Section II the
analytical theory of the 1D MF is presented: the excursion set
approach is reviewed, and the concepts of the {\it collapse radius}
field and of the {\it growth curve} for the objects is
introduced. Section III describes the N-body simulations, based on a
PM code, and the multi-scale algorithm with which the OC regions are
found.  The distinction between points that have a negative Jacobian
and those that are in multi-stream regions is clarified.  Section IV
presents the results of the comparison of theory and simulations.  The
comparison is based both on the collapse radius fields and on the
collapsed objects.  Section V gives a summary of the main results and
the conclusions.

\section{Analytical Theory}

As this 1D analysis is meant to provide tools for addressing the
complex 3D problem, it is useful to consider a 1D universe whose
background geometry is that of a 3D Friedmann model.  This is achieved
by considering a Friedmann universe with purely planar perturbations,
i.e., homogeneous infinitely extended parallel planes interacting via
self-gravitational force.  The evolution of these perturbations is
then 1D.  As the non-linear evolution of perturbations is easily
disentangled from the evolution of the background geometry (see, e.g.,
Monaco 1998\cite{m98}), all the analyses presented in this paper are
restricted to the case of an Einstein-de Sitter model, in which the
geometry is globally flat and the expansion is described by the scale
factor $a(t)\propto t^{2/3}$ (when the universe is dominated by
pressureless matter).  The 3D background density $\bar{\varrho}_{3D}$
is equal to the critical one, $\overline{\varrho}_{3D}
(t)=3H^{2}(t)/8\pi G$, where $G$ is the gravitational constant and
$H(t)=\dot{a}(t)/a(t)$ is the Hubble parameter\cite{p80}.  The density
parameter $\Omega$ is defined as the ratio between the background
density $\bar{\varrho}_{3D}$ and the critical density; in the
Einstein-de Sitter model $\Omega=1$.  The matter field is denoted by
$\varrho(x)$ (mass per unit length), the 1D background density as
$\bar{\varrho}$ (equal to $L^2 \bar{\varrho}_{3D}$, where $L$ is the
length unit), and the density contrast $\delta(x)$ is defined as
$\delta(x)=(\varrho(x)- \bar{\varrho}) /\bar{\varrho}$.  The space
coordinate $x$ is assumed to be comoving with the cosmological
background.

The Fourier transform of the density $\delta$ is denoted by
$\tilde\delta(k)$.  The power spectrum $P(k)$ of the primordial
perturbations of the initial density field is defined as:

\be \label{powspe} P(k)=\mid \tilde\delta_k \mid
^{2}. \label{inspe}\ee

\noindent
It is assumed to be a power-law in $k$:

\be P(k)\propto k^n. \label{power} \ee

\noindent
If the spectral index $n$ is larger than $-1$ the gravitational
evolution of structures follows a hierarchical pattern, smaller
structures collapsing before larger ones.  In the following, the cases
$n=0$ and 1 will be considered.

\subsection{Inverse collapse times in 1D}

The gravitational evolution of a cosmological pressureless fluid can
be described as a map from an initial, quasi-homogeneous configuration
to an evolved one:

\be \mx(\mq,t) = \mq + \ms(\mq,t). \label{map}\ee

\noindent
Here \q\ is the initial, Lagrangian (comoving) coordinate of a fluid
element, \x\ is its final, Eulerian (comoving) location at the time
$t$, and \s\ is called the displacement field.  For small
displacements, the map \s\ can be expressed as a perturbative
series\cite{b94,bch95,c95}, whose first-order term gives the well
known Zel'dovich (1970)\cite{z70} approximation\footnote{ The
first-order term gives the Zel'dovich approximation only for a limited
(but mostly used) set of initial conditions; see Buchert
(1992)\cite{b92}.}:

\be \ms(\mq,t) = -b(t) \nabla_\mq \varphi(\mq). \label{zel} \ee

\noindent 
Here $b(t)$ is the growing mode for linear perturbations\cite{p80},
which is equal to the scale factor $a(t)$ in an Einstein-de Sitter
background cosmology, and \pot\ is the initial peculiar rescaled
gravitational potential, which obeys the equation:

\be \nabla^2_\mq \varphi(\mq) = \delta(\mq,t_i)/b(t_i) \equiv \delta_l.
\label{poisson} \ee

\noindent
The quantity $\delta_l$, which is the initial density contrast
linearly extrapolated to the present time, is called the linear
density.

In 1D, the first-order term of the Lagrangian perturbation series
gives the exact solution of the equations of motion up to OC, where
the map in Eq.~\ref{map} becomes multi-valued.  The density
contrast can be expressed as:

\be 1+\delta(\mq,t)=1/(\partial x/\partial q) = 1/(1-b(t)\delta_l(q)). 
\label{delta} \ee

\noindent
In this case $\delta_l$ takes the role played by the three eigenvalues
$\lambda_i$ of the deformation tensor $S_{a,b}$ in 3D (recall that the
sum of the three $\lambda_i$ gives $\delta_l$), and $J(q,t)=\partial
x/ \partial q$ is the Jacobian determinant of the $\mq\rightarrow\mx$
transformation.

It was shown by Monaco (1995; 1997a,b\cite{m95,m97a,m97b}) that the MF
can be constructed on the basis of the inverse of the collapse `times'
$F\equiv b(t_c)^{-1}$, where the growing mode is used as a time
variable.  The collapse time $t_c$ is defined as the instant at which
the density goes to infinity and is found from the solution of the
following equation:

\be J(q,t_c) = \partial x/\partial q = 0. \label{jac} \ee

\noindent
From Eq.~\ref{delta} it follows that:

\be F(\mq) = \delta_l(\mq). \label{invct} \ee

\noindent
In other words, the 1D dynamical MF is based on the linear density
contrast, which is assumed to be a Gaussian random field.

A very similar inverse collapse time is obtained if it is assumed that
perturbations evolve according to linear theory, and that collapse
takes place if the linear density is greater than a threshold density
$\delta_c$:

\be F_{\rm lin}(q) = \frac{\delta_l(q)}{\delta_c}. 
\label{linear} \ee

\noindent
The fact that the Zel'dovich and linear inverse collapse times are
proportional does not mean that the two descriptions are equivalent.
Indeed, the use of the Zel'dovich approximation allows a clear
definition of collapse as OC, while the definition remains vague in
linear theory.  Moreover, the Zel'dovich approximation fixes the value
of the $\delta_c$ parameter to be one, thus leaving no free
parameters.  Notably, as this method is exact up to OC, the same
result is obtained if 1D `spherical collapse' is used to determine the
parameter $\delta_c$, as usually done in the 3D case.


\subsection{Excursion set approach} 

Eq.~\ref{invct} shows that the dynamical inverse collapse time in 1D
is the Gaussian linear density field.  It follows that the 1D MF can
be constructed by means of the usual PS approach, based on linear
theory.  In more detail, the MF is correctly described by the extended
PS theory, also known as the excursion set approach, proposed by
PH\cite{ph90} and Bond et al. (1991)\cite{bce91}.

The reason for introducing the excursion set approach is the
following.  In the perturbative schemes, and for power spectra that
have power at all scales (like the ones given by Eq.~\ref{power} with
$n>-1$), it is necessary to smooth the initial conditions to remove
the small scale power which would go highly non-linear.  This
procedure is valid under the hypothesis that the dynamics of the
smaller scales that are removed by smoothing do not affect strongly
the behavior of larger scales\cite{bp97}.  Then the dynamical prediction of
collapse is based on the linear density field smoothed with a filter
$W_R(q)$ on a hierarchy of scales $R$.  The `resolution' of the
smoothed field will be denoted in the following both by the smoothing
radius $R$ and by the mass variance $\Lambda$:

\be \Lambda = \langle (\delta_l*W_R)^2 \rangle . \label{lambda} \ee

\noindent
Here the asterisk denotes a convolution.  The mass variance $\Lambda$
is usually denoted by the symbol $\sigma^2$, which is not used here
because it is misleading to use a squared quantity as an independent
variable.  If the power spectrum is a power law, $R$ and $\Lambda$ are
connected as follows: $\Lambda= (R/R_0)^{(1+n)}$, where $n$ is the
spectral index and $R_0$ is the radius corresponding to unity mass
variance.

The excursion set formalism is based on the idea that if a point is
predicted to collapse at the resolution $\Lambda$ (radius $R$), then
it must be considered as collapsed at any larger resolutions (smaller
radii), because the introduction of information from smaller scales
cannot reverse collapse.  The MF problem can be recast into one of
scattering trajectories (in the $F$-$\Lambda$ plane; see Bond et
al. 1991\cite{bce91}) with an absorbing barrier at $F=F_c$, where

\be F_c = 1/b(t). \label{barrier} \ee

The fraction of trajectories absorbed by the barrier at resolutions
smaller than $\Lambda$ (or at radii larger than $R$) gives the
fraction of mass collapsed at that resolution; this quantity is
denoted by $\Omega(<\Lambda)$, and its $\Lambda$-derivative by
$\omega(\Lambda)$\footnote{Note that, in the case of an Einstein-de
Sitter universe, for which the density parameter $\Omega$ is unity,
$\Omega(<\Lambda)$ is the contribution to the cosmological density
coming from collapsed objects at the resolution $<\Lambda$.}.  The
mathematical nature of the problem, and hence its solution, depends
critically on the shape of the window function $W_R$.  If $W_R$ is a
sharp low-pass filter in the Fourier space (sharp $k$-space filter,
hereafter SKS), then the trajectories are random walks, and the
absorbing barrier problem becomes a diffusion problem, which can be
solved through a Fokker-Planck equation. The solution turns out to be
the usual, correctly normalized PS formula

\be \omega(\Lambda)d\Lambda=\frac{F_c}{\sqrt{2\pi\Lambda^3}} 
\exp \left( -\frac{F_c^2}{2\Lambda} \right) d\Lambda.  \label{ps} \ee

If the smoothing is not SKS, the diffusion is subject to a complex
colored noise, and the absorbing barrier problem cannot be solved
exactly.  However, PH have developed a useful approximation to
determine the fraction of collapsed mass; this is given in their 
Eq. 

\bea \omega(\Lambda) = &&\left[\frac{\exp (-1/2\Lambda)}
{\sqrt{8\pi\Lambda^3}} - \int_{-\infty}^{1} P_F(x,\Lambda) dx \right.
\times  \label{ph} \\&& \left.\frac{1}{\pi\Lambda_c\ln 2} \ln
\left(\int_{-\infty}^{1} P_F(x,\Lambda)dx \right)\right] \times
\nonumber \\&& \exp \left[\int_0^\Lambda \ln \left( \int_{-\infty}^{1}
P_F(x,\Lambda')dx\right) \frac{d\Lambda'} {\pi\Lambda_c\ln 2}
\right]. \nonumber \eea

\noindent
Here $P_F$ is the probability distribution function (hereafter PDF) of
$F$, i.e. a Gaussian with zero mean and variance $\Lambda$, and
$\Lambda_c= 2\Lambda\gamma(1-\gamma^2)^{-1/2}$ is a correlation length
for trajectories (the standard spectral measure $\gamma$ is defined in
in PH).

The freedom in the choice of the shape of the window function, and the
different resulting MFs, is one of the most annoying aspects of the
extended PS theory.  Given that filtering is introduced in order to
solve for the dynamics in the mildly non-linear regime, the shape of
the window function must be chosen so as best to reproduce the exact
dynamical evolution of the density field.  This is in the same spirit
as the optimization of the truncated Zel'dovich
approximation\cite{mps94}.  In this sense, Gaussian smoothing is
expected to be preferable.

The transformation from the resolution variable $\Lambda$ or $R$ to
the mass variable $M$ is where the geometry of the collapsed
structures enters into the MF theory.  The knowledge of the fraction
of mass collapsed at a resolution is not enough to determine how the
mass gathers into well-defined clumps. It is reasonable to expect that
the mean size of the clumps is proportional to the smoothing radius,
which is the only typical length present in the problem (as long as
the power spectrum is scale-free or gently curved). Then, an
order-of-magnitude estimate of the mean mass of the clumps is given
by:

\be M = {\rm const} \times\bar{\varrho} R. \label{mass} \ee

\noindent
The exact proportionality constant depends on the shape of the
smoothing window; it is a reasonable choice to keep it as a free
parameter, as in Lacey \& Cole (1994)\cite{lc94}.  With the
transformation given in Eq.~\ref{mass}, it is possible to express the
MF $n(M)$ through the following `golden rule' (as named by Cavaliere,
Colafrancesco \& Scaramella 1991\cite{ccs91}):

\be M n(M) dM = \bar{\varrho} \omega(\Lambda) \left| \frac{d\Lambda}{dM}
\right|  dM. \label{golden} \ee

In practice, a whole distribution of masses forms at a given
resolution; as argued by Monaco (1997b)\cite{m97b}, the actual MF will
be a convolution of Eq.~\ref{ps} or Eq.~\ref{ph} with some
distribution of masses formed at a given radius.  To determine this
distribution one should evaluate the size of the extended patches of
Lagrangian space whose mass elements gather into a collapsed clump.
The excursion set approach cannot be used for this purpose because
mass elements are treated at the 1-point level, neglecting any spatial
correlation between them.  A solution of this geometrical problem thus
requires an extension of the formalism.  We propose a simple
semi-analytical solution.

\subsection{From resolution to mass}

To determine the masses which form at a given resolution, it is useful
to construct a function $R_c(\mq)$\footnote{
It is also possible to construct a function $\Lambda_c(\mq)$.  The
smoothing radius has been used because the resulting plots are easier
to interpret and visualize.},
that gives, for each point \q\ of the Lagrangian space, the largest
collapsing radius at which the point is predicted to collapse (in
other words, at which its trajectory upcrosses the $F_c$ barrier for
the first time).  This function is very useful because it contains in
a compact way all the multi-scale information relevant for the MF
problem.  Fig. 1 shows a typical $R_c(q)$ curve\footnote{
Note that the $R_c(q)$ curve is smooth, the step-like appearance is an
artifact.}
for a realization of a scale-free spectrum with $n=0$ (see Section
III.A for more details).  The intersection of this curve with a line
of constant radius $R$ defines a set of segments; these give the
simply connected regions of points in Lagrangian space that are
collapsed at radii $\geq R$.  Under the reasonable assumption that
each simply-connected collapsed region gathers into a single collapsed
object, the length of each segment (times $\bar{\varrho}$) gives the
mass of the object at the radius $R$.  Due to the solution of the
cloud-in-cloud problem, the excursion sets, and then object masses,
can only grow with decreasing $R$.  It is very important to note that,
although object masses depend on the smoothing radius, for each object
there exists an interval of radius for which the mass does not change
much.  A reasonable definition of the mass of a clump should be
independent of the smoothing radius $R$, which is a spurious quantity
introduced in the theory because we are not able to solve the full
dynamical problem.  Then, the stabilization with $R$ of the mass $M$
of structures allows it to be defined in a meaningful way.

It is possible to construct, for each object, a growth curve $M_i(R;
M_{\rm sat})$, which gives the mass $M$ of the $i$-th object at radius
$R$, saturating at $M_{\rm sat}$.  These curves are not defined for
all $R$ values, as objects appear at some radius and are nested to
other objects at some other smaller radius (this is not a merging
event, as $R$ is not a time variable).  In the limit of infinite
variance, all points collapse, and then all the structures percolate
into one.  This is not worrisome in cases analogous to that shown in
Fig. 1, where the mass of each object is well defined for a
significant range in radius.

The growth curves for objects with mass in an infinitesimal interval
around $M_{\rm sat}$ can be averaged as follows:

\be \langle M_i(R; M_{\rm sat}) \rangle = M_{\rm sat} G(R;M_{\rm sat}). 
\label{mgc} \ee

\noindent
It is assumed that the mean growth curve does not depend on $M_{\rm
sat}$; we have verified the validity of this assumption in our
analysis.  It is convenient to recast the saturating mass in term of a
`saturating radius', which is defined as

\be {\rm const}\times R_{\rm sat} = M_{\rm sat}/\bar{\varrho}.
\label{massa} \ee

\noindent
The value of the constant in this equation is discussed below.

The total mass collapsed at radius $R$ in a large volume $V$ (in
Lagrangian space) is given by a sum over all the $M_i(R)$, the
contributions given at $R$ by all the objects contained in the volume.
The abundance of objects with mass $M_{\rm sat}$ is given by the MF
$n(M)$ evaluated at $M_{\rm sat}$.  Then, as $\varrho V$ is the total
mass contained in $V$, in the limit of infinite volume the fraction of
collapsed mass is

\be \Omega(>R)=\frac{1}{\bar{\varrho}} \int_0^\infty M_{\rm sat}
n(M_{\rm sat}) G(R;R_{\rm sat})dM_{\rm sat}. \label{conva} \ee

It is possible to write Eq.~\ref{conva} as a convolution.  To this
aim, it is convenient to use the mass variance $\Lambda$ as resolution
variable, and to express the mass $M_{\rm sat}$ in terms of a variable
$\Lambda_{\rm sat}=\Lambda(R_{\rm sat})$, where $R_{\rm sat}(M_{\rm
sat})$ is given by Eq.~\ref{massa}.  Under the hypothesis of
mass-independence, the function $G$ is expressible as a function of
$\Lambda/\Lambda_{\rm sat}$.  We define the function
$\tilde{\omega}(\Lambda)$ as the one satisfying the `golden rule'
relation:

\be M(\Lambda) n(M(\Lambda)) dM(\Lambda) = \bar{\varrho}\, \tilde{\omega}
(\Lambda) \left|\frac{dM}{d\Lambda}\right|^{-1} dM(\Lambda).
\label{convb} \ee

\noindent
The $M(\Lambda)$ function is the inverse of $\Lambda_{sat}(M_{sat})$
and the differential fraction of collapsed mass $\omega(\Lambda)$ is

\bea \Lambda \omega(\Lambda)&& d\ln\Lambda =
\label{convc} \\ && \left[ \int_{-\infty}^\infty \Lambda_{\rm sat}
\tilde{\omega}(\Lambda_{\rm sat}) \frac{dG}{d\ln\Lambda}(\ln\Lambda -
\ln\Lambda_{\rm sat})\, d\ln\Lambda_{\rm sat} \right] d\ln\Lambda.
\nonumber \eea

\noindent
In other words, the logarithmic derivative of the fraction of
collapsed mass is the convolution of the $\tilde{\omega}(\Lambda)$
function, which satisfies the golden rule, with the logarithmic
derivative of the mean growth curve for the objects.

The golden rule is recovered in the case in which $G$ is a step
function, and $dG/d\ln\Lambda = \delta^D(\Lambda-\Lambda_{\rm sat})$,
where $\delta^D$ is a Dirac delta function.  This would correspond to
an $R_c$ curve made up by a collection of rectangles, whose heights
are proportional to their widths.  The golden rule can give a fair
approximation to the MF as long as the logarithmic derivative of $G$
is sharply peaked, and as long as the constant in Eq.~\ref{massa}
makes the peak coincide with $\Lambda/\Lambda_{\rm sat}=1$.  Reversing
the argument, the position of the peak of the $dG/d\ln \Lambda$ curve
gives the best constant to use in Eq~\ref{massa}.  As a consequence,
it is not necessary to give the exact constant of proportionality
between radius and mass, as the deconvolution from the differential
growth curve will select the best one; no free parameters are present
in this MF theory.  In the following, the `Gaussian' value
$\sqrt{2\pi}$ will be used both for Gaussian and SKS smoothing.


\section{Simulations}


\subsection{Description of Simulations}

To investigate the dynamics of the cosmic fluid in 1D, we wrote a PM
{\it Particle-Mesh} code\cite{detal80,edfw85} to follow the non-linear
evolution of perturbations.  The PM method of integration has been
preferred over other methods, such as PP ({\it Particle-Particle}),
P$^{3}$M ({\it Particle-Particle/Particle-Mesh}), or TreeCode, because
it is considerably faster, and in one spatial dimension it can reach a
satisfactory resolution by using a large number of grid points.  The
lack of precision of the PM integration for distances smaller than the
grid spacing is not important for our purposes, as we are not
interested in the individual properties of the trajectories of each
fluid element\cite{he81}.

The code integrates the standard Vlasov-Poisson equation for a
collisionless, self gravitating cold fluid in an expanding framework
of coordinates, using a particle sampling of a given initial density
perturbation field. The framework is defined as $r=a(t)x$ , where $r$
is the physical coordinate and $a(t)$ and $x$ are, as before, the
expansion factor and the comoving coordinate.  As discussed at the
beginning of Section II, the expansion has been chosen to follow the
Einstein-De Sitter model for a flat universe.  The one-dimensional
Vlasov-Poisson equations are given and discussed, e.g., in
Doroshkevich et al (1980)\cite{detal80}.  The Poisson equation is
integrated as usual in Fourier space, taking advantage of the FFT
algorithm.  We use a CIC ({\it Cloud-In-Cells}\cite{he81,edfw85})
smoothing algorithm for assigning the mass of the particles to the
grid points and for interpolating the force from the grid points to
the particles.  The velocities and positions of the particles are
evaluated using a time-centered leap-frog scheme, with time-dependent
coefficients that account for the simultaneous expansion of the
comoving framework of coordinates\cite{he81,edfw85}.

The force acting upon the particles is constant, and depends only on
the amount of matter (i.e. on the number of particles) present on the
right and the left of each particle.  In one dimension, the direct
calculation of the forces would be thus exact, so that a PP integrator
could have some advantage with respect to the PM one.  But, if a good
resolution in mass is requested, the PP integrator too time-expensive,
and the PM scheme is preferable.

The initial conditions for all the simulations have been given
consistently with the linear theory of the evolution of perturbations.
The particle sampling of the density perturbation field is obtained
using the Zel'dovich approximation that has been shown to generate a
particle distribution that well reproduces the given power
spectrum\cite{edfw85}.

We tested the code by comparing the analytical linear evolution of a
single-wavenumber perturbation with the numerical integration of the
same perturbation.  Furthermore, we compared the mildly nonlinear
evolution of various density perturbation fields with the results
obtained using the Zel'dovich approximation, which in this case is
exact till OC.  We also included in the code checks of the
conservation of the energy, obtained in comoving coordinates using the
Layzer-Irvine equation\cite{p80}, and of the conservation of linear
momentum.  The code also checks the maximum shift of any particle in
one time step.

As mentioned above, we consider initially scale-free spectra
$P(k)\propto k^{n}$ for the initial density perturbation field
(Eq.~\ref{power}), with $n=0$ and 1.  In these cases, perturbations in
the initial density contrast field are present at all scales of the
simulation, their relative power depending upon the spectral index.
The normalization of a scale-free power spectrum is somewhat
arbitrary; we attach a comoving `physical' length to the simulation
segment, setting it to $L_{box}=1000h^{-1}Mpc$, and normalize in the
`standard' way by defining a smoothed variance, filtered with a
top-hat filter at a scale $R_{\rm TH}=L_{box}/10$, and linearly
evolved to $a_0=1$, as

\be \label{normvar} \Lambda_{\rm TH} (a_0, R_{\rm TH}) =
\frac{1}{2\pi} \int P(k)W^{TH}(kR_{\rm TH})dk=1, \ee

\noindent
where $W^{TH}(kR)$ is the Fourier transform of a 1-dimensional top-hat
filter, and $a_0=1$ is the normalization of the expansion factor.  The
initial time of the simulations has been chosen to correspond to the
epoch $a_{\rm in}$ when the density fluctuations on the Nyquist scale
are equal to the white noise level; therefore the initial epoch of the
simulation are different for the two different spectral indices, while
the differences in output times among the simulations belonging to the
same spectral index depend on the interference among the various
random amplitudes at different $k$ values.  We note that the {\it
physical} initial conditions and intermediate outputs of all the
simulations are equivalent, since they are defined using the variances
and not the simulation times.

For each spectral index, we have performed ten simulations with
different random Fourier phases.  The same seed for the random number
generator has been given to the simulation code when running with
different spectra, so that the two sets of ten simulations and
different spectral indices have the same phases.  The final epoch for
all simulations is $a(t)=1.0$.  One of the simulation for each
spectral index (the first one) has a perturbation amplitude {\it
equal} to the power spectrum, while the others have
Gaussian-distributed amplitudes with {\it variance} given by $P(k)$ at
the scale $k$.  Thus, one of the simulations for each spectral index
is equivalent to an ensemble average over many different realizations
of the initial conditions.  Each simulation has a number of particles
$N_{p}=524288$ and a number of grid-points $N_{gr}=262144$, with a
grid space length of $l=0.0038h^{-1}Mpc$.  We have five order of
magnitude of resolution in length, and four times more in mass.  The
non-conservation of energy is always smaller than 1\%; important bulk
flows are not observed and the total linear momentum is exactly
conserved, at the numerical error level.  The maximum shift of a
particle in a time-step is never greater than $0.8l$.  The number of
time-steps used for each simulation is $N_{st}=19000$.  For each
simulation we have ten outputs, roughly linearly spaced in time,
corresponding to linearly extrapolated variances $\Lambda_{TH}\simeq$
0.0028, 0.0174, 0.0562, 0.135, 0.225, 0.335, 0.468, 0.623, 0.801 and 1
(top-hat smoothing on $1/10$ of the box is again assumed).


\subsection{Negative Jacobian and Multi-Streaming}

A key point in comparing the dynamical MF theory to simulations,
already raised by Monaco (1997a)\cite{m97a}, is that the clump-finding
algorithm must seek those structures that are actually predicted to
form by the theory.  In the present case, the clump-finding algorithm
must look for connected multi-stream regions at a certain scale.  In
fact, a large density contrast is a consequence, due to the continuity
function, of OC and multi streaming.  Standard clump-finding
algorithms based on percolation or on overdensities, such as the
friend-of-friends (hereafter FOF) one or others seeking density peaks,
may not be the best choice for finding multi-stream regions, because
they typically find them in a biased way.  This is the case in 3D;
more compact, roundish clumps are typically preferred over filamentary
or sheet-like structures (even by FOF if the linking length is chosen
so as to pick up the virialized clumps), making such algorithms
unsuitable for the comparison with a theoretical prediction based on
OC and multi streaming.  Moreover, standard clump-finding algorithms
have a free parameter, like the FOF linking length or a threshold
density, while our theoretical predictions do not.

When searching for multi-stream regions, a difficulty immediately
appears.  Fig. 2 shows the \x\ vs \q\ plot of a simple Gaussian
perturbation: the triple-valued region, in which triplets of different
elements \q\ get to the same Eulerian point \x, is the multi-stream
region.  The slope of the $\mx(\mq)$ curve gives the Jacobian
determinant of the transformation.  The `OC' condition of having a
negative Jacobian, Eq.~\ref{jac}, is then satisfied for those points
\q\ where the slope of the curve is negative.  This defines a segment
in Lagrangian space which is shorter than the whole multi-stream
region.  In the following, these regions will be called NJ regions,
while the whole multi-stream regions will be called MS regions.

The origin of the problem is that mass elements which have a negative
Jacobian according to the Zel'dovich approximation are subject to OC
and get shocked only after crossing the structure. However, the
Zel'dovich approximation is not a good tool for describing the
behavior of mass elements subject to multi-streaming: after entering
an MS region, the mass element starts to interact highly non-linearly
and non-locally with the other mass elements, without waiting a whole
crossing time to be shocked.  Then, the Zel'dovich approximation is
not able to recover the whole MS region, which is nevertheless
dynamically relevant.  On the other hand, it can be a fair
approximation to say that the NJ collapse condition finds mass
elements that have approximately first-crossed an MS region.  If MS
regions were infinitely thin, such as in the adhesion theory, then
there would be no difference between them and NJ ones.  So this
problem is connected to the finite width of MS regions.  As long as
the density around caustics shows some kind of universal
profile\cite{sz89,nfw95}, one expects a tight correlation between the
masses of MS- and NJ-based objects.

To find the NJ and MS regions in the simulation, the Eulerian
positions of particles have been evaluated on a mesh grid in
Lagrangian space, made up by 65536 grid points, through a CIC
interpolation.  The resulting $\mx(\mq)$ curve has been smoothed with
a hierarchy of Gaussian filters of decreasing radius $R$.  NJ regions
have been identified with connected regions (segments) characterized
by a negative slope of the $\mx(\mq)$ curve, while MS regions have
been identified with connected regions of points in triple-valued
regions.  This algorithm is fast, easy to implement and does not
contain free parameters.  Its multi-scale nature is analogous to the
multi-scale nature of the MF problem; as a consequence it is possible
to construct $R_c(q)$ curves (like the one shown in Fig. 1) for the NJ
or MS regions in the simulations by recording the largest smoothing
radius $R_c$ at which the particle at grid point $q$ enters an NJ or
MS region.  As in Section II.C, it is possible to construct growth
curves for the $R$-dependent objects, defined as segments in the
intersection of the $R_c$ curve with a line of constant $R$, and the
mass of these objects is well defined as long as the growth curves for
the objects saturate at some value.

The definition of structure proposed here is different from more
standard ones, although the continuity equation assures that both MS
and NJ regions are tightly related to high-density clumps.  The
importance of the algorithm described above relies on its connection
with the important dynamical concepts of NJ and MS: the NJ and MS
structures are important regardless how they compare to clumps defined
in more standard ways.  Moreover, the performance of standard
clump-finding algorithms has not been thoroughly tested in the 1D
case, which is remarkably different from 3D in many regards.  As a
consequence, we will focus mainly on the NJ and MS structures, as
their analysis allows us to deepen the dynamical problem and develop
many useful techniques extendable to the 3D case.  Anyway, a
comparison of NJ and MS structures to FOF clumps will be shown in
Section IV.D for completeness.


\section{Comparison of Theories with Simulations}

The $R_c(\mq)$ curves, as the one shown in Fig. 1, have been
constructed from the analytical predictions of collapse, both with
Gaussian (GAU) and SKS smoothing, applied to the initial conditions of
the simulations described in Section III.A, resampled on a grid in
Lagrangian space with 65536 grid points.  In the SKS case, it is
necessary to follow all the modes of the box to sample the $F$
trajectories in all the points; to keep the analysis feasible, only
the first 10000 modes have been considered.  Analogous $R_c(q)$
curves have been found by seeking NJ and MS regions in the
simulations, as described in Section III.B.

Fig. 3 shows the $R_c(\mq)$ curves for the third output of the first
simulation, both with $n=0$ and $n=1$ (the third output is chosen as
it is the first one in which the differential growth curve is stable;
at the NJ and MS level, it shows more structure than the following
ones).  Only one tenth of the box is shown.  In particular, Figs. 3a
and 3c show a comparison of GAU and NJ curves, Figs. 3b and 3d a
comparison of the SKS and MS curves.  Some important points can be
appreciated from a visual inspection of Fig. 3:

\begin{enumerate}
\item 
The theoretical predictions based on Gaussian smoothing are able to
reproduce accurately the NJ regions in the simulations (Figs. 3a and
3c).  A cross check with Figs. 3b and 3d shows that SKS structures are
systematically larger that GAU ones.  It is confirmed that Gaussian
smoothing optimizes the dynamical prediction of NJ.
\item
Surprisingly, SKS predictions reproduce, although not with great
accuracy, the MS regions (Figs. 3b and 3d).  This `coincidence' is due
to the fact that the SKS smoothing overestimates the size of GAU
collapsed regions nearly as much as the MS regions overestimate the
size of the NJ regions.
\item
SKS and MS structures tend to include more than one GAU or NJ
structure.
\item
All structures, especially NJ and GAU ones, show a general tendency
toward saturation.
\item
The agreement of theory and simulations is better for $n=0$ than for
$n=1$.  This is expected, as $n=1$ gives more small-scale power, which
provides, through highly non-linear dynamics, noise in the predictions
relative to larger scales.
\end{enumerate}

In the following subsections all the conclusions just drawn are
statistically quantified.  Section IV.A gives a global analysis of the
$R_c(\mq)$ curves, Section IV.B gives object-by-object comparisons,
Section IV.C shows the resulting MFs and Section IV.D shows a comparison
with FOF objects.


\subsection{Global Analysis}

Three statistics are used to quantify the agreement between different
$R_c(q)$ curves: the coincidence statistic, which is based on counting
points for which two different predictions agree or disagree, the
Pearson and the Spearman correlation coefficients.  The comparison of
two $R_c$ curves is done by cutting the curves to progressively larger
radii, so as to test the effect of small scale objects.

The coincidence statistic is defined as follows.  We count the number
of points in the Lagrangian grid for which neither, one or both curves
predict collapse.  The number of points for which neither of the two
curves predicts collapse results of minor interest, as at large
$R$-values it is dominated by the large number of points which have
not collapsed.  The statistic chosen for the comparison is the ratio
between the number of points for which both curves predict collapse
and the number of points for which at least one of the two give
collapse.  In formal terms, the coincidence statistic is defined as
follows:

\be C = \frac{P(R_{c1}(\mq)>R_{cut}\ {\rm AND}\ R_{c2}(\mq)>R_{cut})}
{P(R_{c1}(\mq)>R_{cut}\ {\rm OR}\ R_{c2}(\mq)>R_{cut})}. 
\label{coinc} \ee

\noindent
Here $R_{c1}$ and $R_{c2}$ are the two curves being compared, $P({\rm
event})$ denotes the probability of the event within the brackets, and
$R_{cut}$ is the cut in smoothing radius.  The coefficient $C$ takes
values from 0 (anti-coincidence) to 1 (perfect coincidence).

The correlation of two $R_c$ curves is also measured by means of two
standard statistical indicators, the Pearson linear coefficient $r_p$
and the Spearman non-parametric rank coefficient $r_s$\cite{ptv92}.

Fig. 4 shows the three correlation coefficients $C$, $r_p$ and $r_s$
for the pairs of curves shown in Fig. 3 (the results for the other
simulations are similar).  The agreement between GAU and NJ curves is
good at all scales, with correlation coefficients larger than 0.8 in
the $n=0$ case; the agreement is still present but worse in the $n=1$
case.  The agreement between SKS and MS curves is not as good, but a
strong correlation is still present at all scales, with correlation
coefficients always in excess of 0.6.  Note that the increase of the
coincidence coefficient $C$ in Figs. 4b and 4d is artificial: the SKS
predictions are not pushed to very small radii, and this lack of small
structures mimics a real lack of small MS objects.


\subsection{Object-by-object analysis}

An object is defined, at a given smoothing scale $R$, as a segment
defined by the intersection of the $R_c(\mq)$ curve with a line of
constant $R$.  The growth curves $G$ for objects are constructed as
follows.  Two segments at $R$ and $R-\Delta R$ are considered part of
the same growth curve when the one at $R-\Delta R$ includes the one at
$R$ (segments at $R$ cannot be partially included in segments at
$R-\Delta R$).  Two growth curves are nested if they end up in the
same segment; in this case the growth curve corresponding to the
smaller object is stopped, while the larger one takes the mass of both
structures and continues growing.  A growth curve is considered as
saturated when its mass changes by not more than 5\% for a change of
$R$ by a factor 1.3.  It has been verified that the catalogues of
objects and their statistical properties do not change significantly
when the saturation conditions are slightly changed.  The saturating
mass of the object $M_{\rm sat}$ is calculated as the average mass
within the central half of the $R$-interval of saturation.

Some objects fail to saturate; these are in general small objects,
whose growth curves have not saturated at the smallest radius
considered in the analysis or have been nested to a larger object.  In
this case the object is assigned the mass at its smallest radius.
Such objects are considered in the analysis, but not to construct the
final growth curve.

After saturation, growth curves continue to grow by accreting small or
tiny objects.  As all the objects end up percolating at some small
scale, it is very important to define the mass of objects at their
first saturation.  However, the mass which is further acquired by the
objects is not assigned to any smaller clump.  This highlights a
problem with the definition of isolated clumps; any uncertainty in the
definition of larger clumps is at the expenses of smaller ones.  The
choice made here of neglecting the mass further accreted by the object
can lead to a decrease of the number of small mass objects.  However,
there is no reason to believe that assigning this mass to small
objects would be a better choice.

The comparison of two different object catalogues is done by
associating the objects in one catalogue with the ones in the other
catalogue if their associated segments at the saturation mass
intersect.  An object of a catalogue can thus be associated with more
than one object in the other catalogue.  In fact, most of these
associations are between objects of quite different sizes.  It is
interesting to consider the average number of objects in catalogue B
associated with a single object in catalogue A, which will be denoted
by $\bar{n}_{AB}$, and the same average number limited to those
objects whose mass is at least 10\% of the mass of the object they are
associated with, denoted by $\bar{n}^{10\%}_{AB}$.  We have also
considered the fraction of mass of the largest and second largest
objects in catalogue B, in terms of the mass of the object in
catalogue A with which they are associated, which will be denoted by
$M2/M1_{AB}$ (if there is not a second largest object, its mass has
been set to 0). This quantity is important; if it is small, the MF is
likely not to be severely affected by a disagreement at the
object-by-object level.  Note that all these quantities change if the
A and B catalogues are commuted; then, the pairs of catalogues have
been compared in both senses.

Fig. 5 shows these three quantities for pairs of catalogues of all the
objects found in all the outputs of all the simulations.  Masses are
rescaled to the value of $M_*$, and the results are shown in four bins
of $M/M_*$.  The comparisons shown are again GAU vs NJ and SKS vs MS,
in both the $n=0$ and $n=1$ cases.  The results obtained by inverting
the two catalogues considered are also shown, except for the
$M2/M1_{AB}$ indicator, which does not change much on inverting the
catalogues.

The comparison of GAU and NJ catalogues is shown in Figs.  5a and 5c.
Although NJ regions tend to include many small GAU structures,
especially for $n=1$, the number of `large' associated objects is not
much different from one.  The $M2/M1_{AB}$ index is always smaller
than 0.1, indicating that the second associated objects are negligible
in terms of mass.  As before, the two catalogues agree more in the
$n=0$ case.

The comparison of the SKS and MS catalogues, shown in Figs. 5b and 5d,
reveals a good object-by-object agreement, although not as good as
that of GAU and NJ objects.  In particular, SKS objects tend to
include many MS objects at large masses, but again the mass of the
second associated object is almost negligible.

Fig. 6 shows a one-to-one comparison of the masses of associated
objects, in units of $M_*$.  All the objects of the third output of
all the simulations are shown.  The masses refer to the pair of
largest associated objects; then, only a subset of all the objects is
shown.  The masses show a tight correlation in all cases, which is
tighter for $n=0$.  The correlation is almost unbiased, even though
objects found from simulations tend to be more massive than predicted.

In conclusion, the catalogues extracted from simulations with the NJ
or MS algorithms reveal a good agreement with the ones predicted with
GAU or SKS filters.  It is then demonstrated that in 1D the dynamical
MF theory agrees with the simulations on an object-by-object basis,
once the clump-finding algorithm is designed to find exactly what is
sought for.  The agreement worsens if the spectral index $n$ is
larger, so that more small-scale power is present in the spectrum.


\subsection{Mass Functions}

In Section II.C it was shown that to obtain the MF, the fraction of
collapsed mass $\Omega(<\Lambda)$ must be deconvolved from the mean
growth curve of objects.  The mean growth curve is constructed by
averaging the single growth curves described in the last subsection.
The independent variable for the growth curves is given by the
logarithm of the ratio between the two variances $\Lambda$ and
$\Lambda_{\rm sat}$, corresponding respectively to the smoothing
radius $R$ and the radius $R_{\rm sat}$ which corresponds to the total
mass $M_{\rm sat}$ of the object through Eq.~\ref{massa}.  The average
is performed over all saturated objects in all the ten simulations, at
fixed output time.  It has been verified that the resulting mean
growth curve does not depend on the output time.  In more detail, the
first two outputs show somewhat different growth curves, as the
objects are still small and their growth curves are not sampled well
enough.  The results shown and used in the following are those
relating again to the third output.

Figs. 7 shows the mean integral and differential growth curves for GAU
and SKS objects, for the two spectra considered.  As discussed in
Section IV.B, objects continue their growth after saturation.
Consequently, the mean growth curves do not saturate exactly at unity.
The differential growth curve is truncated so as to be properly
normalized.  The differential growth curves show a prominent peak.
The peak position is different for the GAU and SKS curves, showing
that the SKS curve requires a larger constant in Eq.~\ref{massa}.
The peaking of the differential growth curves confirms that the golden
rule can be a good approximation for estimating the masses of the
objects.  The position of the peak depends on the spectrum, for the
GAU filter it decreases with $n$ increasing (the consequences of this
will be discussed in the Conclusions).  The $n=0$ curves are more
sharply peaked than the $n=1$ curves.  The SKS curves are much more
noisy, especially at small variances.  This is a consequence of the
nasty oscillations of the SKS filter in real space.

Fig. 7 shows also the $\omega(\Lambda)$ curves, given in
Eq.~\ref{ps} and Eq.~\ref{ph}, and those deconvolved from the
differential growth curve ($\tilde{\omega}(\Lambda_{\rm sat})$,
Eq.~\ref{convc}).  The deconvolution is performed numerically; this
is a delicate and unstable calculation, such that the large-mass part
of the mass function can be spoiled by numerical oscillations,
especially in the SKS case, where the small-variance tail of the
differential growth curve is very noisy.  However, the numerical
deconvolution turns out to be satisfactory in the GAU case

The excursion set formalism is able to predict, through
Eqs.~\ref{ps} and \ref{ph}, the integral ($\Omega(<\Lambda)$) or
differential ($\omega(\Lambda)$) fraction of mass collapsed at a
variance $\Lambda$.  This prediction is exact for SKS filters, and a
good approximation for Gaussian filters, and hence we expect to find a
very good agreement between the analytical and semi-analytical curves
(i.e. obtained from the GAU and SKS $R_c$ curves).  The accuracy of
the agreement between analytical and semi-analytical predictions gives
then a measure of the fluctuations present in the initial conditions
of the simulation especially at large scales, due to the power-law
nature of the power spectra used.  These fluctuations are enhanced in
the simulation by the non-linear coupling of the first two modes of
the box.  The effect of this non-linear coupling is also seen in
Fig. 3, where the $R_c$ curves taken from the simulations are
systematically higher than the analytical curves.  In other
realizations, where the first two modes interact destructively, the
numerical curves tend to be lower than the analytical ones.  As long
as this enhancement does not influence the width of the spikes in
Fig. 3, neither the catalogues of objects nor their MF is going to be
strongly affected by this effect, which is however revealed by
analyzing the fraction of collapsed mass $\Omega(<\Lambda)$.

We have found that at least ten simulations are necessary to average
out these large-scale fluctuations.  Large masses, or small variances,
remain affected by the fluctuations as they are determined by a few
objects per simulation.  Fig. 8 shows the comparison of the analytical
(PS and PH), semi-analytical (GAU and SKS) and numerical (NJ and MS)
$\Omega(<\Lambda)$ curves, for $n=0$ and $n=1$; the third output of
all simulations is used.  The analytical and semi-analytical curves
agree well, as expected.  Some disagreement is present at small
variance, due to shot noise, and in the comparison between the GAU and
PH curves, due to the approximate nature of the PH analytical
prediction when Gaussian filters are used.  The $\Omega(<\Lambda)$
curves from the simulations show strong fluctuations at small
variance, and a tendency to predict more collapsed mass, which is
consistent with the slightly larger masses of the objects from the
simulations, visible in Fig. 6.  The agreement with the analytical and
semi-analytical curves is good at moderate and large variance for
$n=0$, but is not satisfactory for $n=1$.  From a more detailed
analysis of the simulations, which is not shown for reasons of
brevity, it appears that this behavior is caused by the non-linear
coupling of small and large-scale modes, which is stronger with $n=1$
than with $n=0$, and causes the large-scale fluctuations to influence
small scales.

In the scale-free cases analyzed in this paper, the MFs at different
outputs can be rescaled by showing the quantity $M_*n(M/M_*)dM/M_*$.
Due to the resolution and statistical limits, every output samples a
different part of the mass function, so that a good range in mass can
be achieved by using all the outputs.  Constructing a mass function
from a weighted average of all the outputs is not correct in
principle, as the outputs are not independent.  However, showing the
results of many different outputs is quite confusing.  For this reason
we perform an average, weighted by the number of objects in the bin,
of the MFs from different outputs, with the {\it caveat} that this
procedure is motivated mainly by graphical reasons.  Fig. 9 shows the
GAU, SKS, NJ and MS MFs of all the outputs, together with the averaged
ones.  The stability of the averaging procedure is apparent.

The MFs shown in Fig. 9 are compared in Fig. 10. The analytical
predictions of PS and PH are shown, both those obtained with the
standard golden rule (with Gaussian mass) and those obtained with the
deconvolution procedure of Section II.C.  The following points can be
noted:

\begin{enumerate}
\item
The GAU and NJ mass functions are in very good agreement.  Consistent
with what was found above, the NJ curve tends to give more massive
object, and is slightly flatter (for $n=0$) at smaller masses.
\item
The deconvolved PH curve improves significantly the agreement between
the analytical and semi-analytical GAU curves, confirming the validity
of the deconvolution procedure of Section II.C.  The residual
disagreement at large masses is probably motivated by the fact that
the large mass part of the GAU MF is mainly determined by the first
output, for which the growth curves are not well sampled.  The
small-mass slope is fairly well reproduced, implying that the mass
neglected by the clump-finding algorithm does not influence
significantly the MF in the mass range tested.
\item
The agreement between the SKS and MS MFs is very good.  The
deconvolution does not greatly improve the agreement between the
analytical and semi-analytical predictions, because of the difficulty
in determining a good mean growth curve for the SKS objects.
\item
Both the SKS and MS MFs tend to cut off at small masses.  This is an
artifact in the SKS case, as small smoothing radii are not considered,
but it is real in the MS case, as large MS regions tend to include the
smaller ones.
\item
Consistent with what was found above, the results with $n=1$ are worse
than those with $n=0$.
\end{enumerate}

It is worth noting that no free parameters are involved in the
comparison of the different MFs.


\subsection{Friends-of-friends groups}

The clump-finding algorithm developed in this paper has been
constructed to reproduce the NJ and MS regions in Lagrangian space.
Moreover, this algorithm is based on the saturation condition for the
growth curves, which, as shown in Section IV.C, is not exactly
achieved, and this can lead to a decrease of small-mass objects.  In
this sense the algorithm, although physically motivated, may be
considered not ideal for general application.

Standard clump-finding algorithms are typically applied to the evolved
configuration of a simulation in Eulerian space.  Such algorithms are
designed to select simply connected clumps (like the
friends-of-friends one, hereafter FOF\cite{fof}) or high-density peaks
(DENMAX, SKID\cite{denmax}, HOP\cite{hop} or SO\cite{so} ones).  These
algorithms have been developed and tested for analyzing 3D
simulations, but generally have not been used in 1D.  Therefore, the
use of such algorithms in our case is not guaranteed to lead to
reliable results.  We have nevertheless decided to analyze our
simulations with the standard and most frequently used FOF algorithm,
in order to compare its performance to our OC-based one.

With the FOF algorithm, groups are defined as those sets of particles
whose distance from at least another member of the group is smaller
than a given linking length.  In 3D, it is usual to set this linking
length to 0.2 times the average interparticle distance; in this way
the selected groups have an overdensity of at least $\sim$180, which
corresponds to the density contrast of a spherical virialized clump.
In 1D this argument does not hold, and therefore we have left the
linking length as a free parameter, fixing it as that which gives a
mass function which best resembles the other numerical ones.  A
compromise value of 1.5 times the mean interparticle distance has been
used.  In this way the FOF MF reproduces the large-mass end MS MF for
$n=0$ and the NJ MF for $n=1$.  We have checked that moderate
variations of the linking length lead to similar results.  Remarkably,
the value of the linking length used is larger than that used in 3D.
This can be explained by the fact that 1D clumps, having a more
limited number of degrees of freedom, do not relax and virialize as
easily as in 3D.  As a consequence their internal distribution has a
larger amount of substructure, made up of the caustic fronts which are
generated by multi-streaming.  Then, a larger linking length is
necessary not to fragment the multi-streaming regions.  This
highlights the fact that 1D FOF clumps are remarkably different from
3D ones.

A basic prediction of the mass function theory is that structures
correspond to simply connected regions in the Lagrangian space.  For
an appropriate comparison of the FOF mass function with the other ones
presented above, in which structures are defined as segments in
$q$-space, it is necessary to verify that this hypothesis holds for
the FOF clumps (which are by construction connected only in the
$x$-space).  To check this, we have calculated, for each FOF clump
found from the output of a simulation, the Lagrangian distance between
two particles (starting from the median particle) which contain a
fixed fraction of mass.  If the clump is simply connected (it is a
segment in the $q$-space), and if the mass is expressed in units of
comoving length (with $\bar{\varrho}=1$), the two quantities (distance
and fraction of mass) would be equal.  An estimate of the connectivity
of the clumps is then given by the fraction of mass (in unit of
Lagrangian length) of the clump included in the Lagrangian distance
defined above, averaged over all the objects of a simulation.  In
fig. 11 we show this curve for the third output of the first
realization of $n=0$; other outputs give very similar results even for
$n=1$.  If the hypothesis of simple connection held exactly, then the
function would coincide with the bisector line shown in the Figure.
It is apparent that the deviation from the simply connected hypothesis
is not severe, especially in the inner parts of the clump, while the
borders, which contain no more than 10\% of the mass, are more
fragmented.

However, the object-by-object comparison of the FOF and GAU or SKS
catalogues of objects reveals a poor agreement, as shown in Fig. 12.
The figure shows the results of performing the same analyses as
described in Section IV.B and in Fig. 5 and 6, only for the case $n=0$.
The results for $n=1$ are very similar.  Figs. 12a and b show that
significantly more than one object of a catalogue is associated with
each object of the other catalogue, and that the second largest object
is always significant in terms of mass.  Figs. 12c and d show that the
correlation between the largest companions is still significant but
with a considerable scatter, much larger than the previous cases. The
comparison between NJ and MS groups and FOF clumps is very similar,
and is not shown, as we are not interested at this level in comparing
different 1D clump-finding algorithms.

Fig. 13 shows the MFs of FOF groups compared with the ones shown in
Fig. 10, for $n=0$ and $n=1$.  Only the third output of the first
realization is shown, as it is enough to make the point.  In the $n=0$
case the FOF MF is similar to the MS and SKS ones, but it is more
peaked at $M=M_*$, and the small-mass slope is shallower.  The peak at
$M=M_*$ is even more evident in the $n=1$ case, where the FOF MF cuts
off similarly to the GAU and NJ ones.  It is noteworthy that, by
tuning the linking length, one could reproduce the large-mass cutoff
of any of the other MFs, without changing much the (dis)agreement at
the object-by-object level.  Moreover, at variance with the 3D case,
the linking length should change with the spectrum to reproduce the
cutoff of either the GAU or the SKS MFs with different spectra.

As a conclusion, the analytical mass function does not reproduce well
the FOF mass function of the simulation, both statistically and on an
object-by-object base.  In our opinion, the difference between this
result and the standard 3D case arises predominantly from the
different behavior of the FOF algorithm in 1D relatively to the
widely tested 3D case.


\section{Summary and Conclusions}

In this paper we have shown a complete analysis, performed with
analytical, semi-analytical and numerical techniques, of the
cosmological MF problem with 1D gravity.  When gravitational collapse
is identified with the first crossing of the orbits, the dynamical
solution of the MF problem, analogous to that reviewed by Monaco
(1998)\cite{m98} for the 3D case, coincides formally with that of
linear theory, but with no free parameters.  Then, the MF is well
described by the usual extended Press \& Schechter
theory\cite{ps74,ph90,bce91}.  The determination of the size of
collapsed regions is performed by constructing (with semi-analytical
techniques) a mean growth curve for the objects, which gives the
distribution of masses that form at a given smoothing radius.  The
analytical MF is then determined by a deconvolution of the well known
formulae of Press \& Schechter\cite{ps74} and Peacock \&
Heavens\cite{ph90} from the mean growth curve.

To test the predictions of the MF theory, two sets of ten simulations
of flat expanding universes have been performed.  The two sets differ
only in the power spectrum of primordial perturbations, which is
assumed to be a power-law with index $n=0$ or 1.  The simulations have
been analyzed by searching for the regions in the Lagrangian space
which have a negative Jacobian (NJ) or are in the multi-stream regime
(MS).  The difference between NJ and MS regions arises as an effect of
the finite size of MS regions.  Notably, this kind of analysis is
intrinsically multi-scale.  The analytical predictions of collapse
have been applied to the initial conditions of the simulations, in
order to have semi-analytical predictions for the objects found in the
simulations, and then to test the recovery of the MF object by object.
To obtain such predictions, both Gaussian and sharp $k$-space
smoothing filters have been used.

To characterize in a compact way the multi-scale collapse of the
density field, we have introduced the function $R_c(q)$, which gives
the largest smoothing radius at which the point $q$ in Lagrangian
space is predicted or `observed' (in the simulation) to collapse.  The
$R_c$ curves have been constructed both for the semi-analytical
predictions of collapse and for the NJ and MS regions found in the
simulations.  The semi-analytical predictions of collapse have then
been tested by comparing the $R_c$ curves; the agreement has been
quantified by using several correlation indices.  Catalogues of
objects have been defined starting from the $R_c$ curves, and
exploiting the tendency of simply connected collapsed regions in
Lagrangian space to be stable for a range of smoothing radii.  In this
way the definition of object becomes effectively independent of the
resolution at which the density field is filtered.  The
semi-analytical and simulated catalogues of objects have been compared
both on an object-by-object basis, using several indices to quantify
the agreement, and statistically, by comparing the resulting MFs.

The overall result is that the dynamical MF theory in 1D is able to
reproduce the collapsed objects in the simulation both statistically
and object by object, once the clump-finding algorithm applied to the
outputs of the simulation is designed so as to find the objects which
are in fact predicted, i.e. the NJ or MS regions.  In more detail,
Gaussian smoothing allows an optimal reconstruction of NJ regions, as
anticipated by Monaco (1997a,b)\cite{m97a,m97b}, while the objects
found with SKS smoothing resemble the MS regions.  The good agreement
is reached without tuning any free parameter.  As expected, the
agreement is better if the spectral index is $n=0$, because in the
$n=1$ case the small-scale power present enhances the effect of highly
non-linear dynamics, which is simply removed in the analytical and
semi-analytical predictions.  If the standard friends-of-friends
algorithm is used to define the collapsed clumps in the simulations,
then the agreement is much worse, especially at the object-by-object
level.  However, the performance of the friends-of-friends algorithm
has not been thoroughly tested in 1D.

The analyses presented in this paper have shown that the solution of
the mass function problem in 1D is important for understanding the
dynamics of gravitational collapse, as its degree of complexity is
high enough to be interesting, but low enough to be fully manageable.
The results of this paper are then useful to address the 3D case.
However, the 3D problem presents several important complications, so
that the relevance of the simple 1D study is not obvious.

Firstly, while in 1D the Zel'dovich approximation is exact up to OC,
in 3D further-order terms in the Lagrangian perturbation theory must
be considered.  However, it was shown by Monaco (1997a)\cite{m97a}
that the Lagrangian series converges in predicting the collapse of
mass elements, provided that it is not too slow.  Pushing the
calculations to third order is enough to achieve a satisfactory degree
of precision in the dynamical prediction of collapse, at least for the
large-mass part of the mass function.

Secondly, in 3D the mass elements are subject to collapse along three
different directions, so that it is in principle possible to define
either a ``first'' collapse along the first axis, or a collapse along
the second or the third axis.  First-axis collapse does not separate
fully-virialized clumps and pancake-like or filamentary transients, so
that third-axis collapse could be a better choice for isolating
virialized halos\cite{bm96,ls98}.  The difference between the two
cases is well illustrated in the collapse of a spherical peak which
has a decreasing density profile\cite{m98}: all mass elements except
the central one are subject to cylindrical symmetry, and never
collapse along the third axis.  Then, collapse along the third axis is
able to pick up the seeds for structure formation, and works well if
applied, for instance, to the peaks of the density field (as done,
e.g., by Bond \& Myers 1996\cite{bm96}), but it is not a good choice
for recovering all the collapsing mass.  The fact that Lee \&
Shandarin (1998)\cite{ls98} need a fudge factor of 12.5 to achieve a
good normalization for their mass function can be seen as a
consequence of this fact.

Thirdly, while the collapsed field in 1D remains separated in
simply-connected regions down to large variance, the 3D collapsed
field percolates at moderate variance.  In other words, the excursion
sets of the $R_c({\bf q})$ field in 3D soon become geometrically and
topologically complex.  There is need of a further criterion to
fragment the collapsed medium into objects; such criterion could for
instance be based on the peaks of the $R_c({\bf q})$ field.  Only
after this criterion is defined, it is possible to construct the
growing curve for the objects.  This is related to the important (and
not trivial) problem of assessing the relation between regions in the
multi-stream regime (either NJ or MS) and the clumps found by standard
algorithms.  We stress again that, whatever the result, the NJ or MS
regions are interesting by themselves, because of their dynamical
meaning.

An analysis of 3D simulation is in progress, and some preliminary
results have already been presented in Monaco (1999)\cite{m99}.  The
$R_c({\bf q})$ curves have been calculated, using Gaussian smoothing,
for a realization of a Gaussian field on a cubic grid, used as initial
condition for a numerical P$^3$M simulation.  The ${\bf x}({\bf q})$
map from the simulation has been smoothed and differentiated, so as to
calculate the Jacobian determinant in each point and then find the NJ
regions.  The agreement between predictions and NJ regions is good as
expected.  Further analyses show an encouraging similarity between NJ
regions and FOF groups, mapped in the Lagrangian space, at variance
with the 1D case.  As discussed in the text, the NJ condition peaks up
the collapsed points only after their first crossing of the structure,
and this helps in excluding the less relaxed regions.  The
fragmentation of the collapsed medium cannot be uniquely
defined\cite{m99}, and this is related to the fact that there is not a
unique clump-finding algorithm.

Finally, we want to stress the importance of the concept of the mean
growth curve, which allows the solution of a long-standing problem of
the MF, namely that of a realistic (and simple!)  treatment of the
geometry and topology of collapsed structures in the MF theory.  In
section IV.C (see Fig. 7) it is suggested that the relation between
resolution and mass can change with the spectral index, as when more
small-scale power is present, the excursion sets are more fragmented.
If confirmed in the 3D case, this would explain the change of the
$\delta_c$ parameter detected in the simulations of Governato et
al. (1998)\cite{gbq98}.

\acknowledgments

The authors thank Annalisa Bracco, Paolo Catelan, Vince Eke, Antonello
Provenzale and Tom Theuns for many discussions, and the anonymous
referee for his comments.  P.M.  thanks George Efstathiou for his
advice and support, and the Trans-Edit Group of Milano for
hospitality.  P.M.  has been supported by the EC TMR Marie Curie grant
ERB FMB ICT961709.


\begin{figure}
\centerline{
\psfig{figure=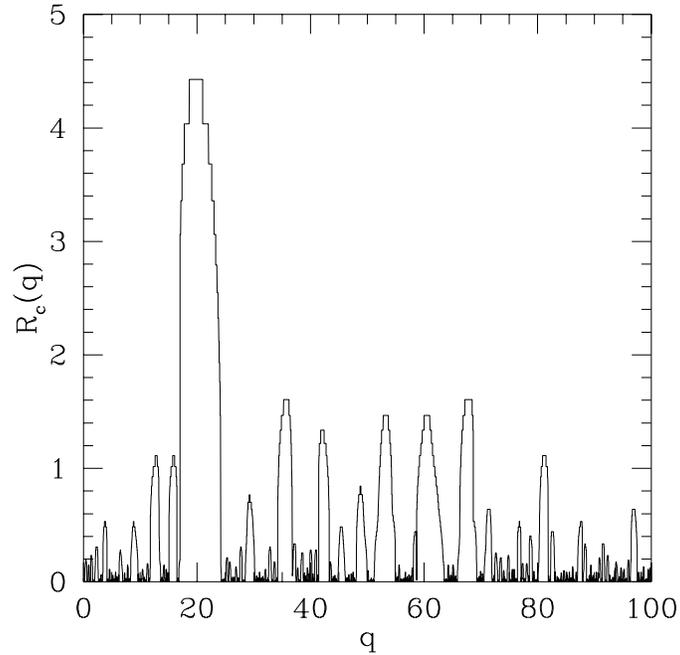,width=9cm}
}
\caption{$R_c(\mq)$, largest collapse radius as a function of \q, for
a realization of an $n=0$ spectrum (Section III).}
\end{figure}

\begin{figure}
\centerline{\psfig{figure=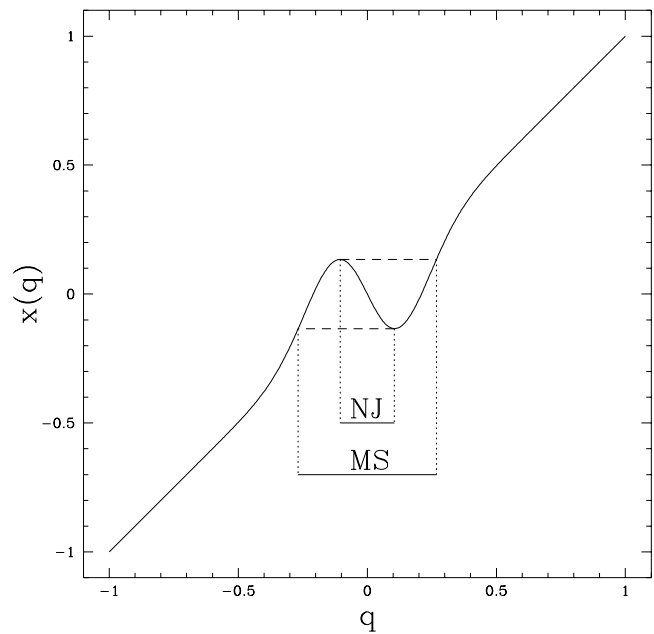,width=9cm}}
\caption{Illustration of the difference between NJ and MS zones.}
\end{figure}

\begin{figure}
\centerline{ \psfig{figure=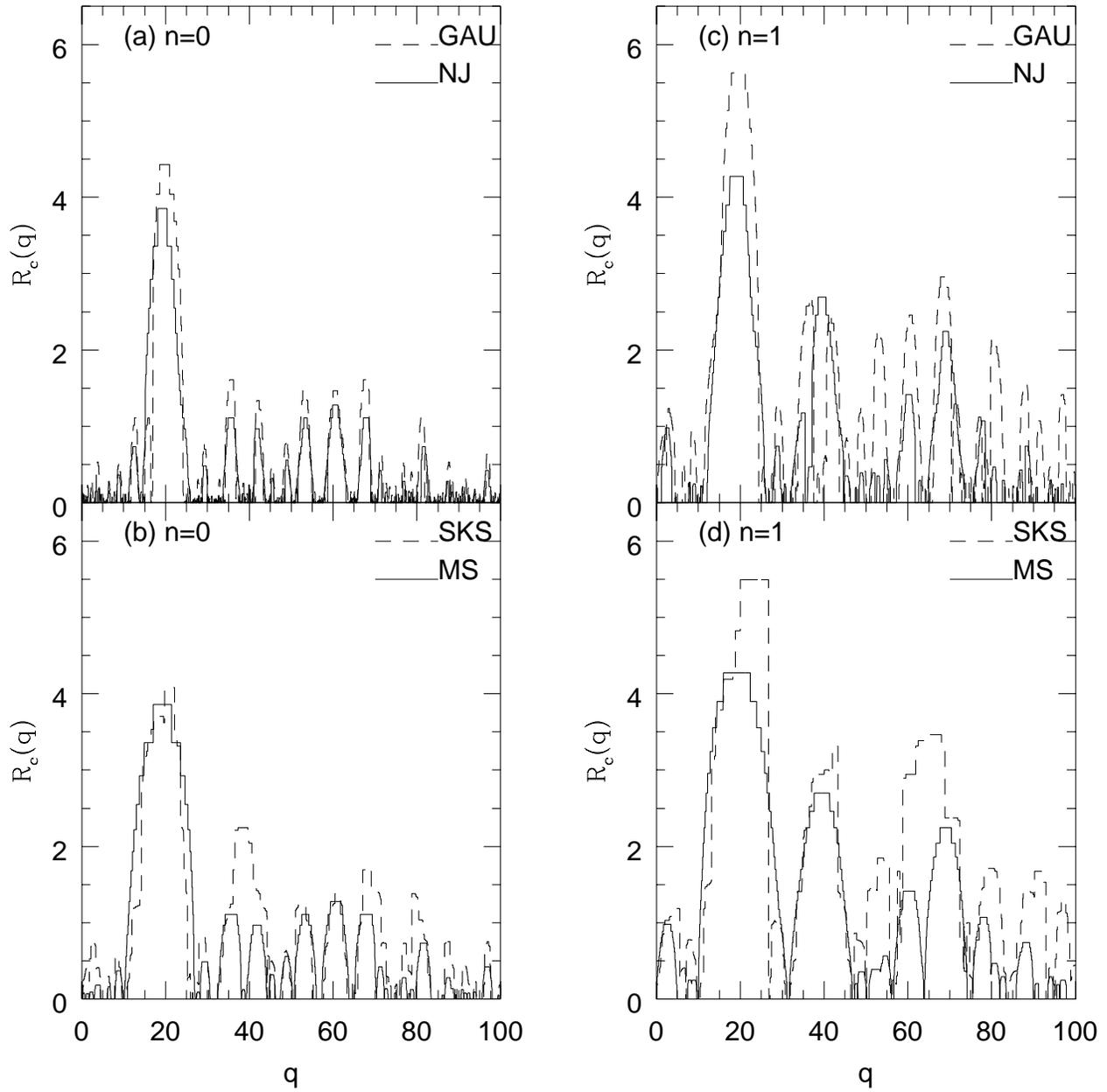,width=18cm} }
\caption{$R_c(\mq)$ curves for different collapse definitions and 
different spectra}
\end{figure}

\begin{figure}
\centerline{
\psfig{figure=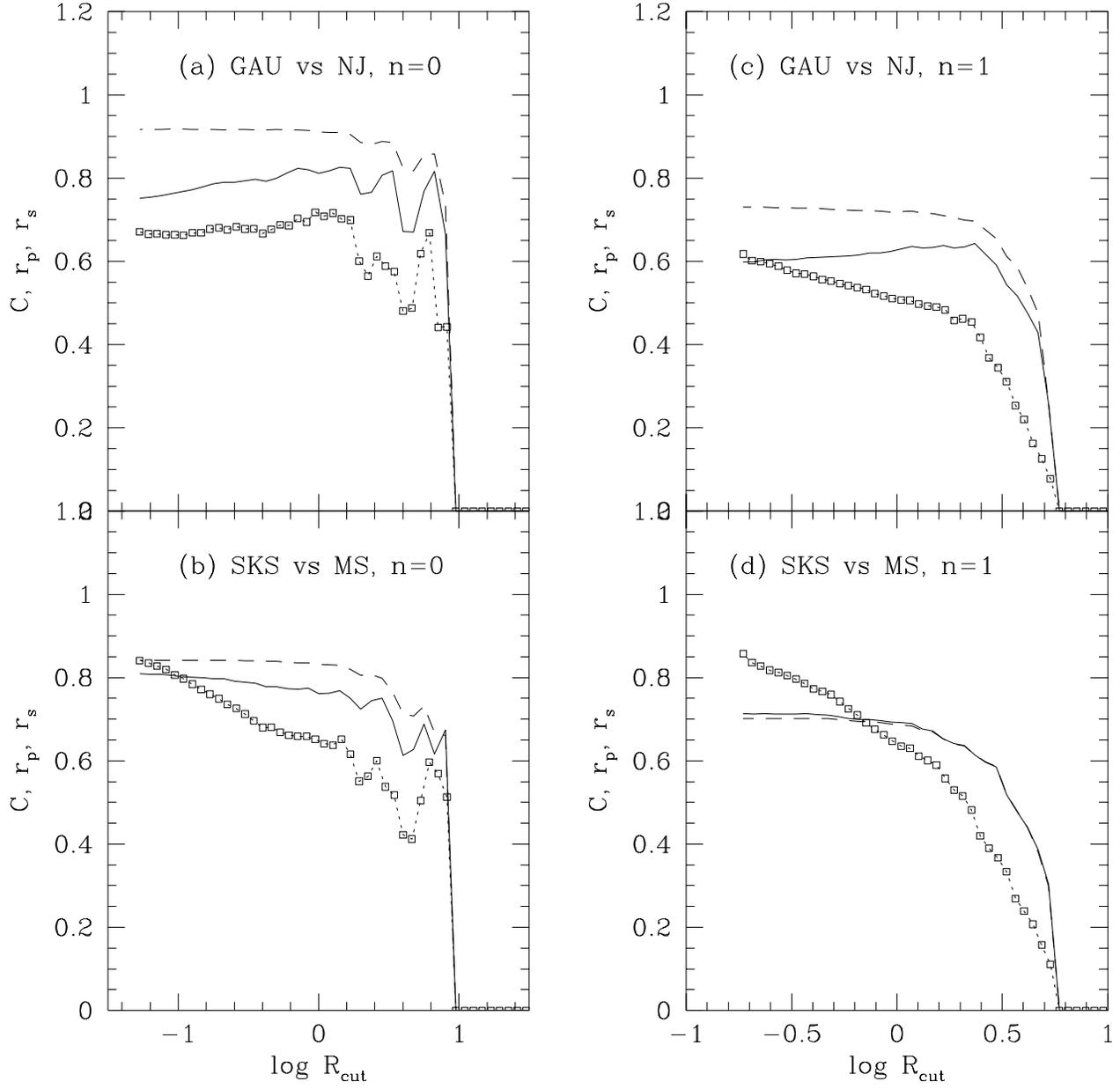,width=18cm}
}
\caption{Global correlations of different $R_c(\mq)$ curves.
Dashed line: Spearman coefficient $r_s$.  Continuous line:
Pearson coefficient $r_p$. Dotted line with square points:
coincidence statistics.}
\end{figure}

\begin{figure}
\centerline{
\psfig{figure=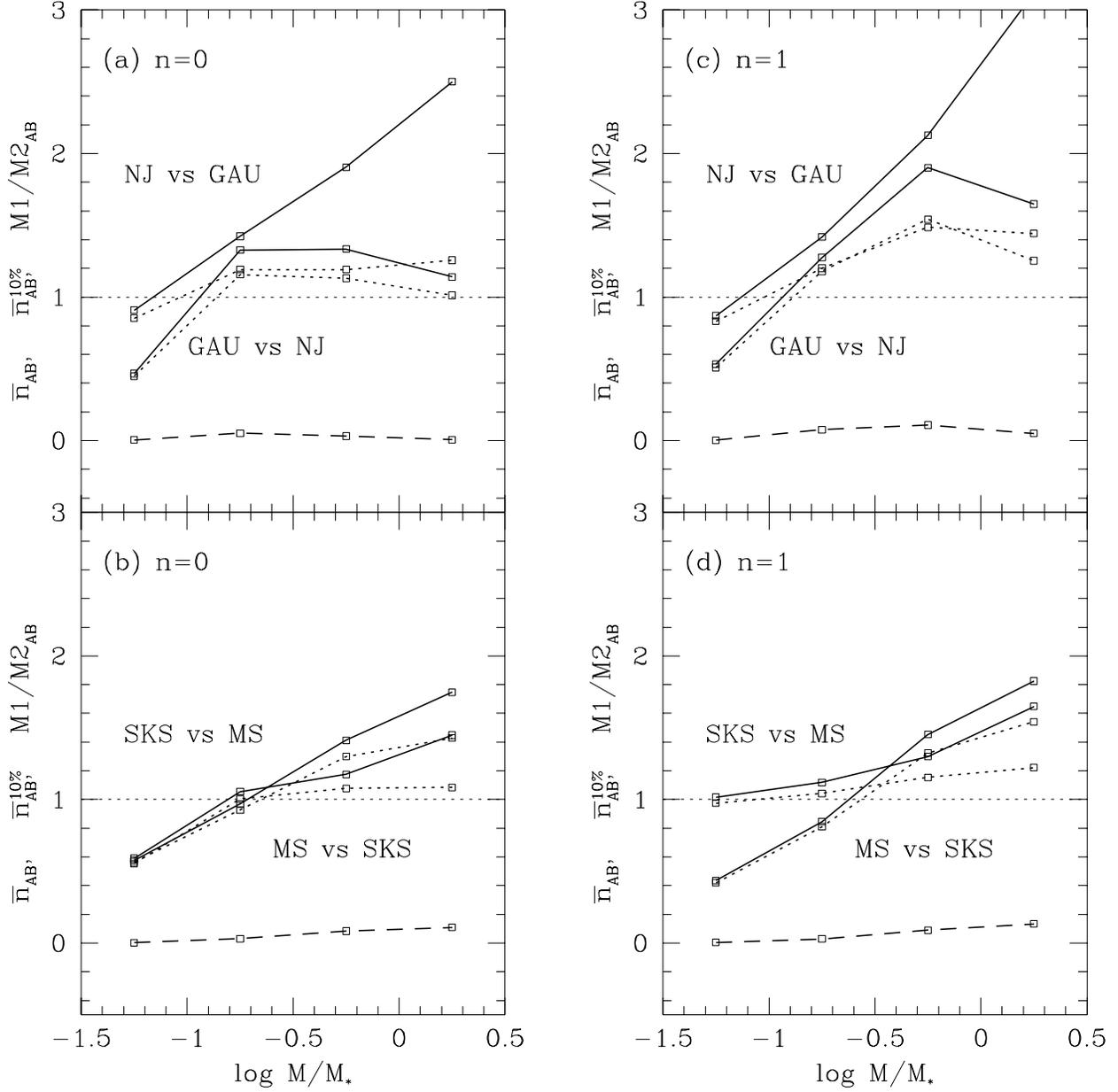,width=18cm}
}
\caption{Comparison of different catalogues of objects.  Continuous
line: $\bar{n}_{AB}$.  Dotted line: $\bar{n}^{10\%}_{AB}$.  Dashed
line: $M2/M1_{AB}$}
\end{figure}

\begin{figure}
\centerline{
\psfig{figure=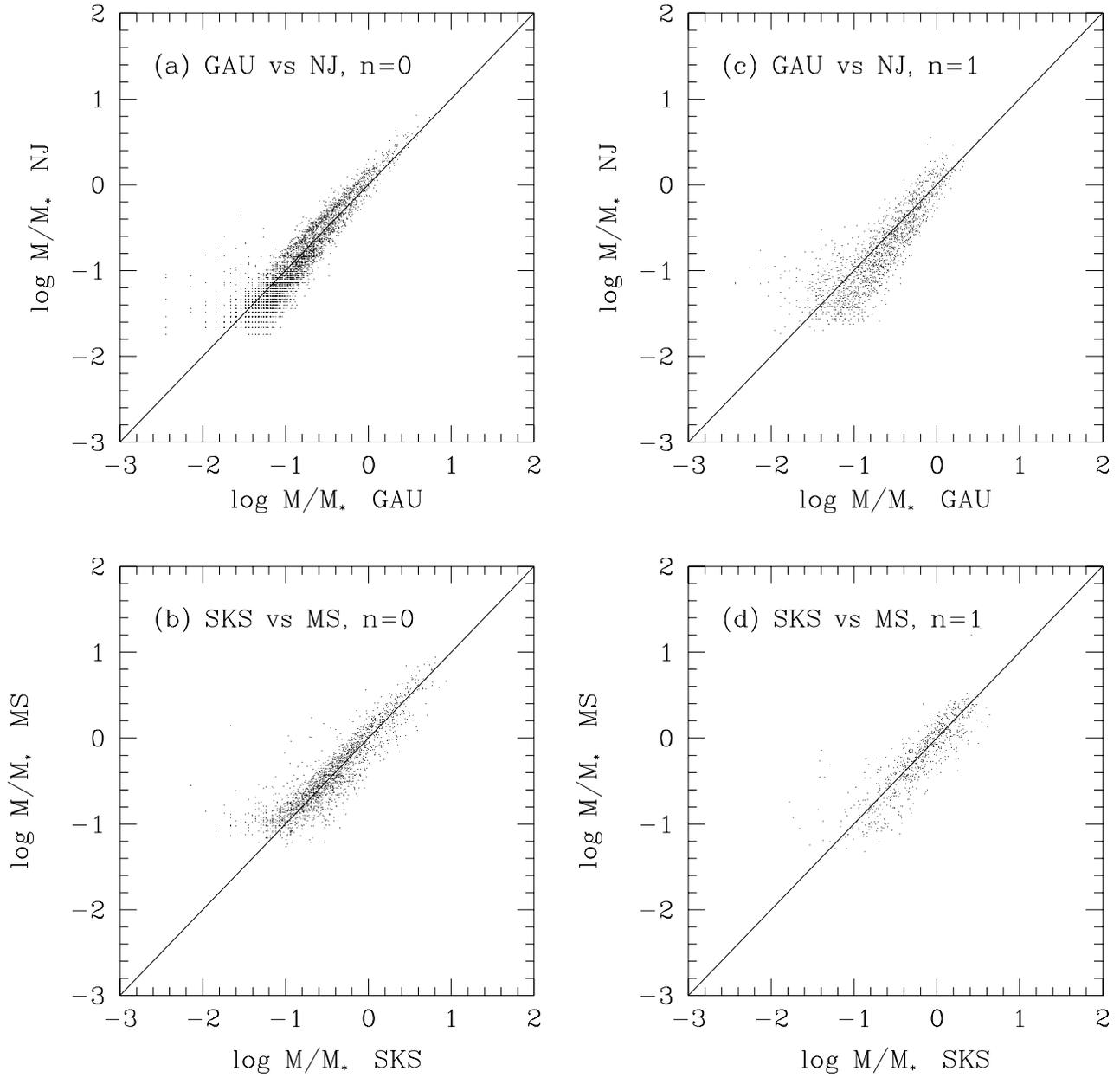,width=18cm}
}
\caption{Comparison of object masses.}
\end{figure}

\begin{figure}
\centerline{
\psfig{figure=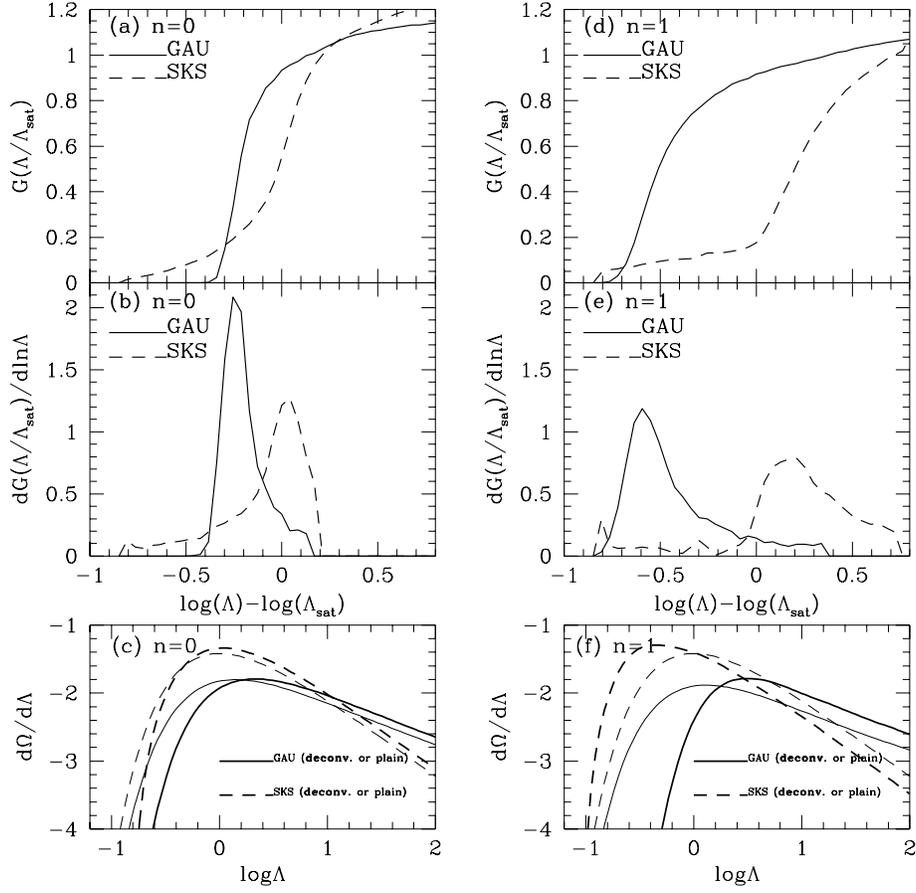,width=13cm}
}
\caption{Growth integral and differential curves; deconvolved
$\omega(\Lambda)$ functions.}
\end{figure}

\begin{figure}
\centerline{
\psfig{figure=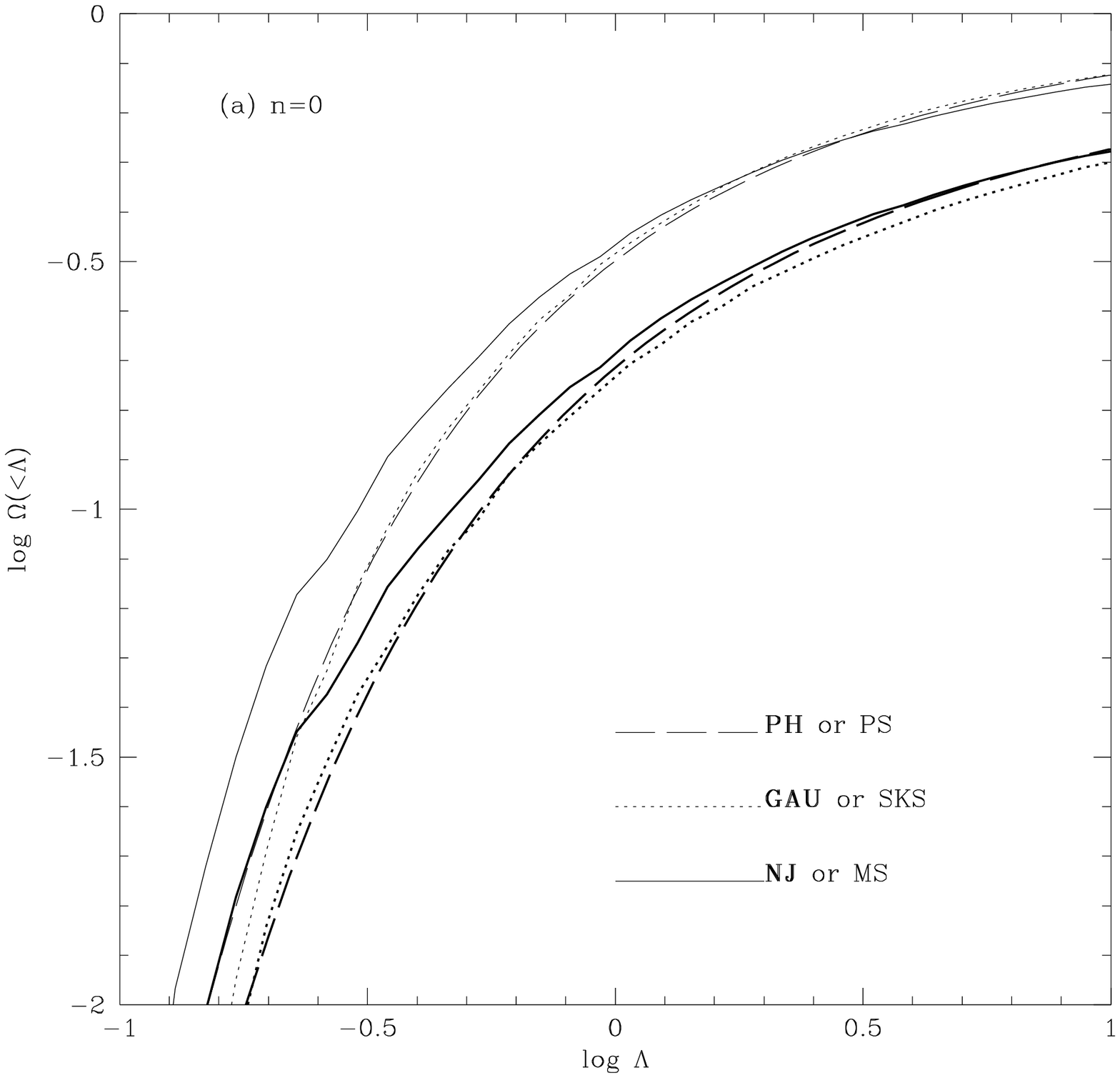,width=8cm}
\psfig{figure=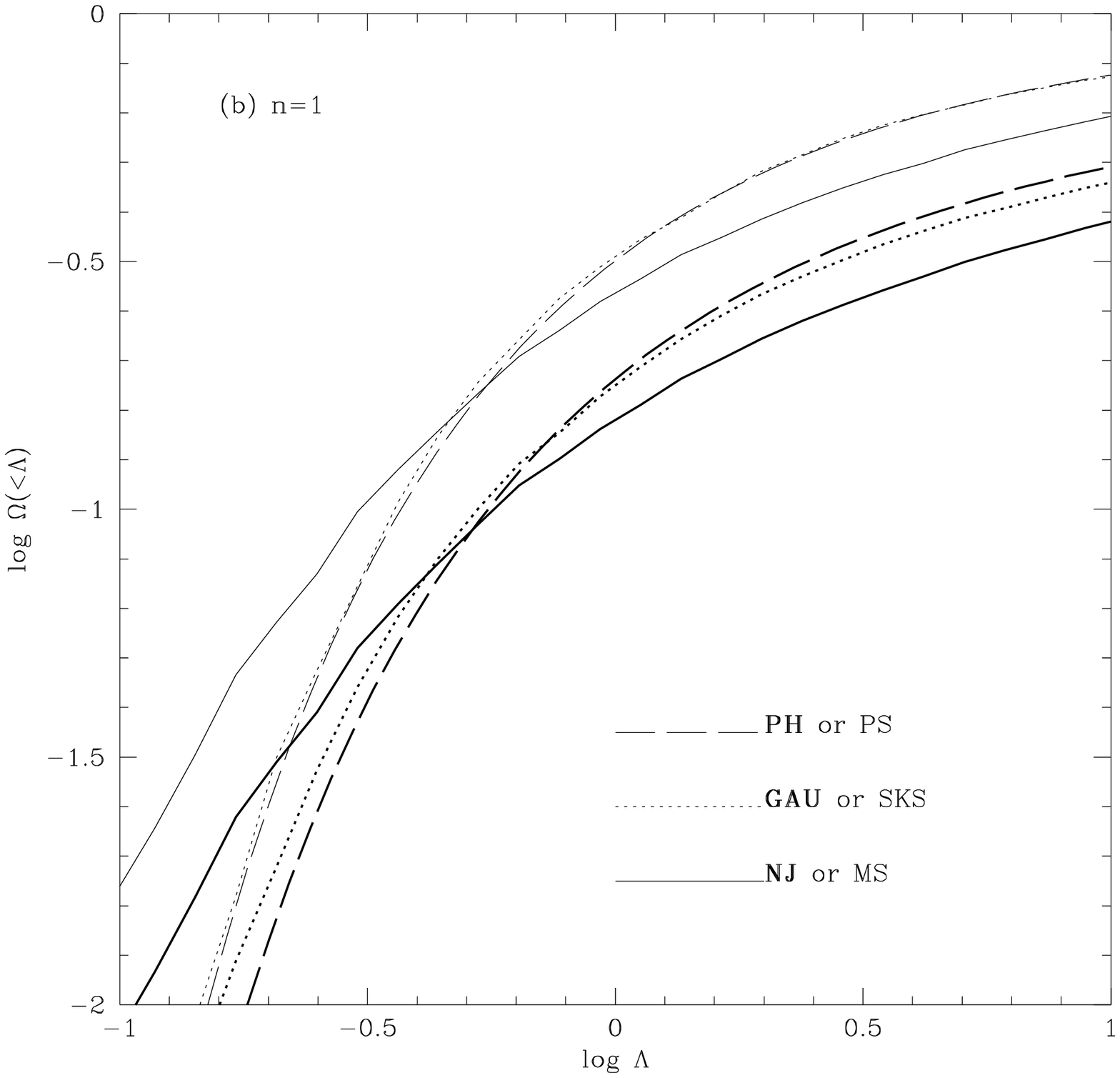,width=8cm} }
\caption{Fraction of collapsed mass.}
\end{figure}

\begin{figure}
\centerline{
\psfig{figure=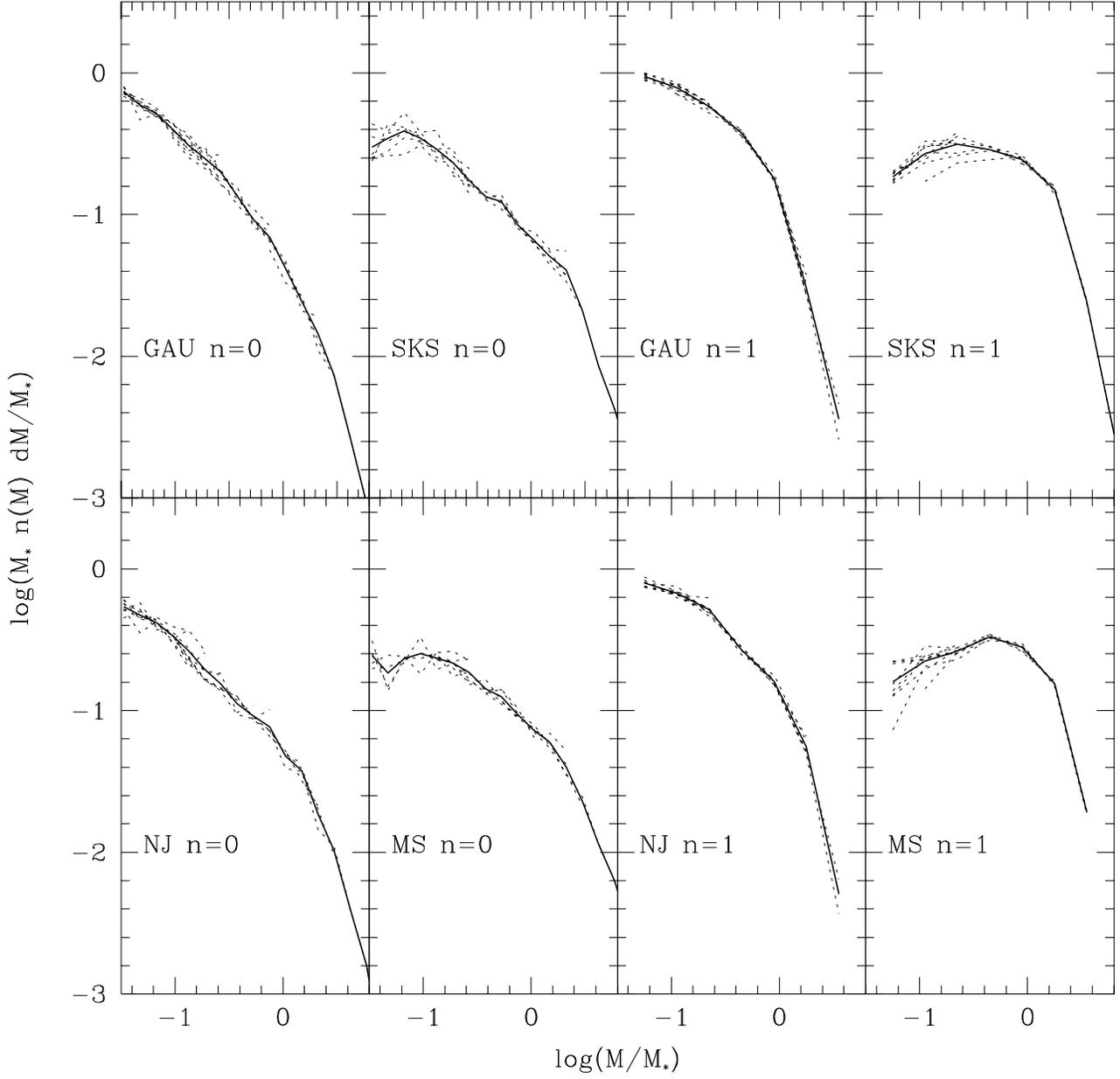,width=18cm}
}
\caption{Mass functions from different outputs: dashed lines represent
single outputs, solid lines show the average.}
\end{figure}

\begin{figure}
\centerline{
\psfig{figure=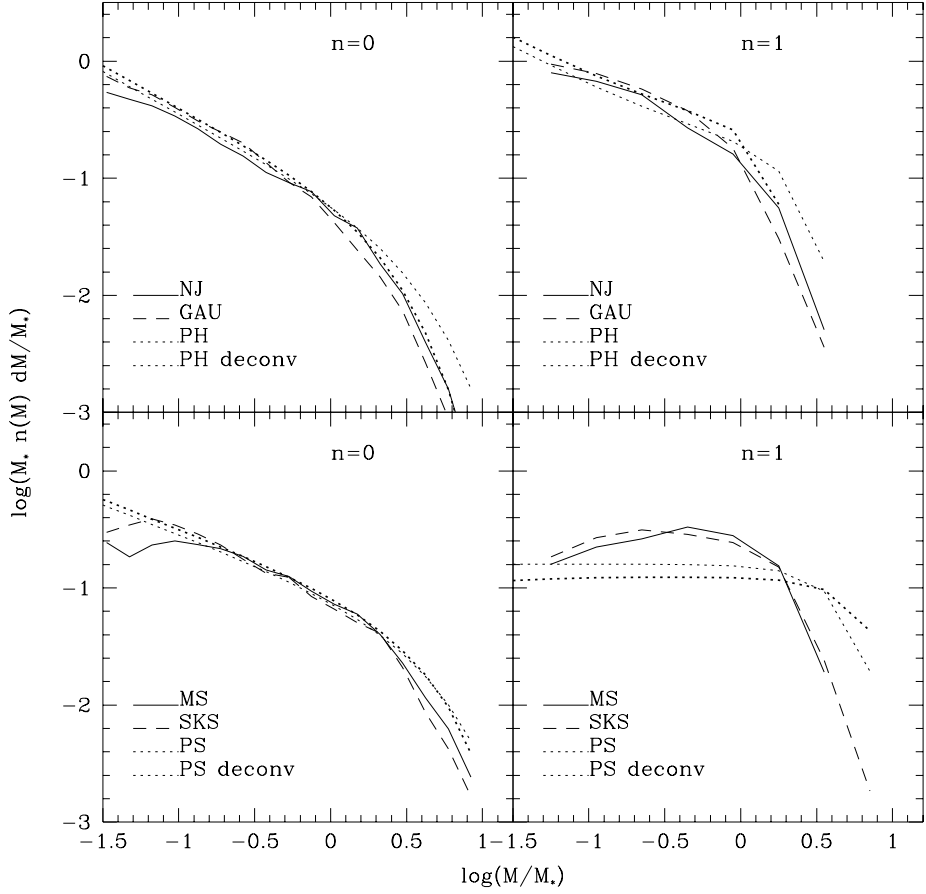,width=13cm}
}
\caption{Comparison of different mass functions.}
\end{figure}

\begin{figure}
\centerline{
\psfig{figure=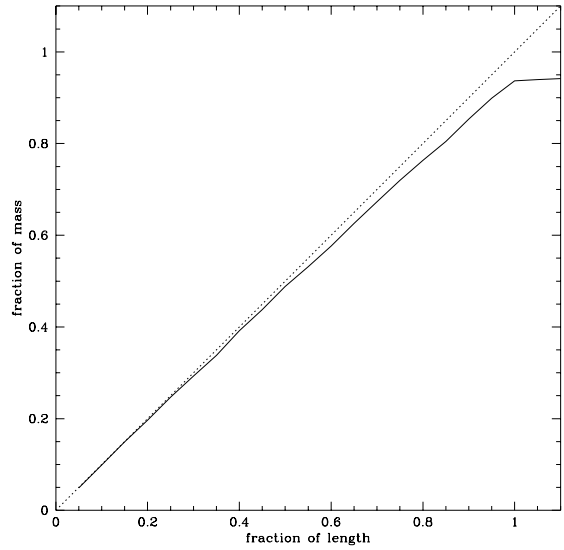,width=8cm}
}
\caption{Test of connectivity of FOF groups in Lagrangian space.}
\end{figure}

\begin{figure}
\centerline{
\psfig{figure=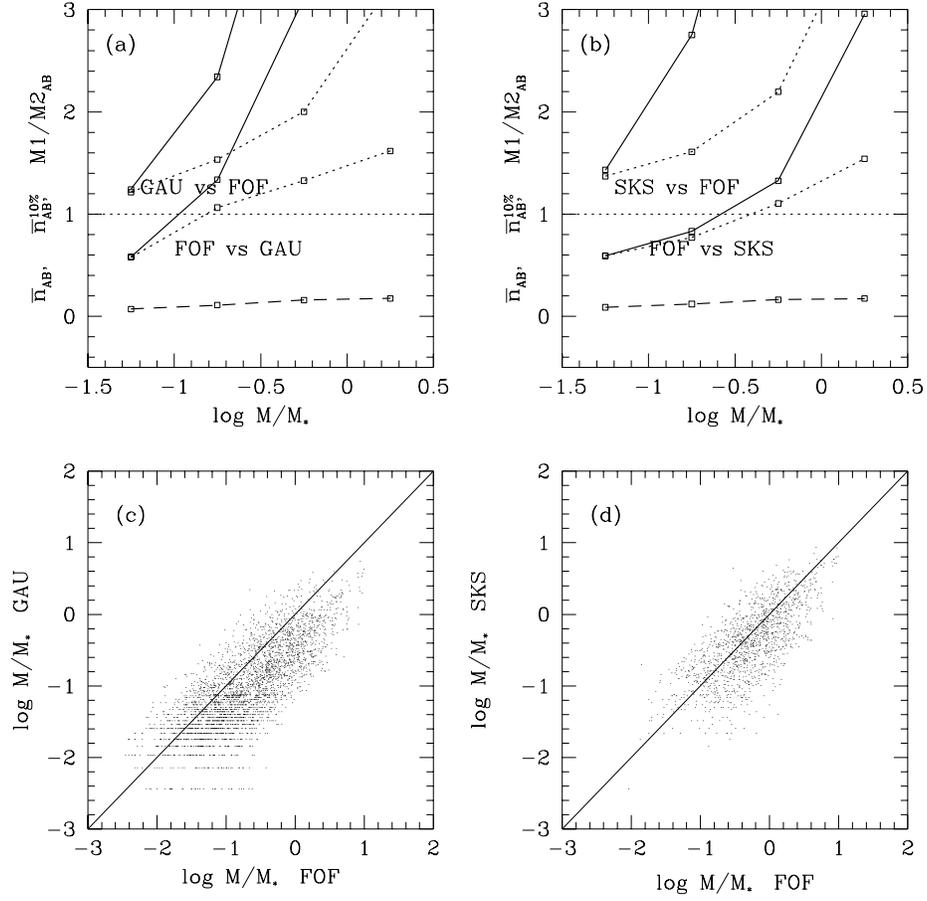,width=13cm}
}
\caption{Object-by-object comparison of FOF objects catalogues with
GAU and SKS ones. In panels (a) and (b): continuous line:
$\bar{n}_{AB}$.  Dotted line: $\bar{n}^{10\%}_{AB}$.  Dashed line:
$M2/M1_{AB}$.  Panels (c) and (d) show the correlation of the masses
of associated objects.}
\end{figure}

\begin{figure}
\centerline{
\psfig{figure=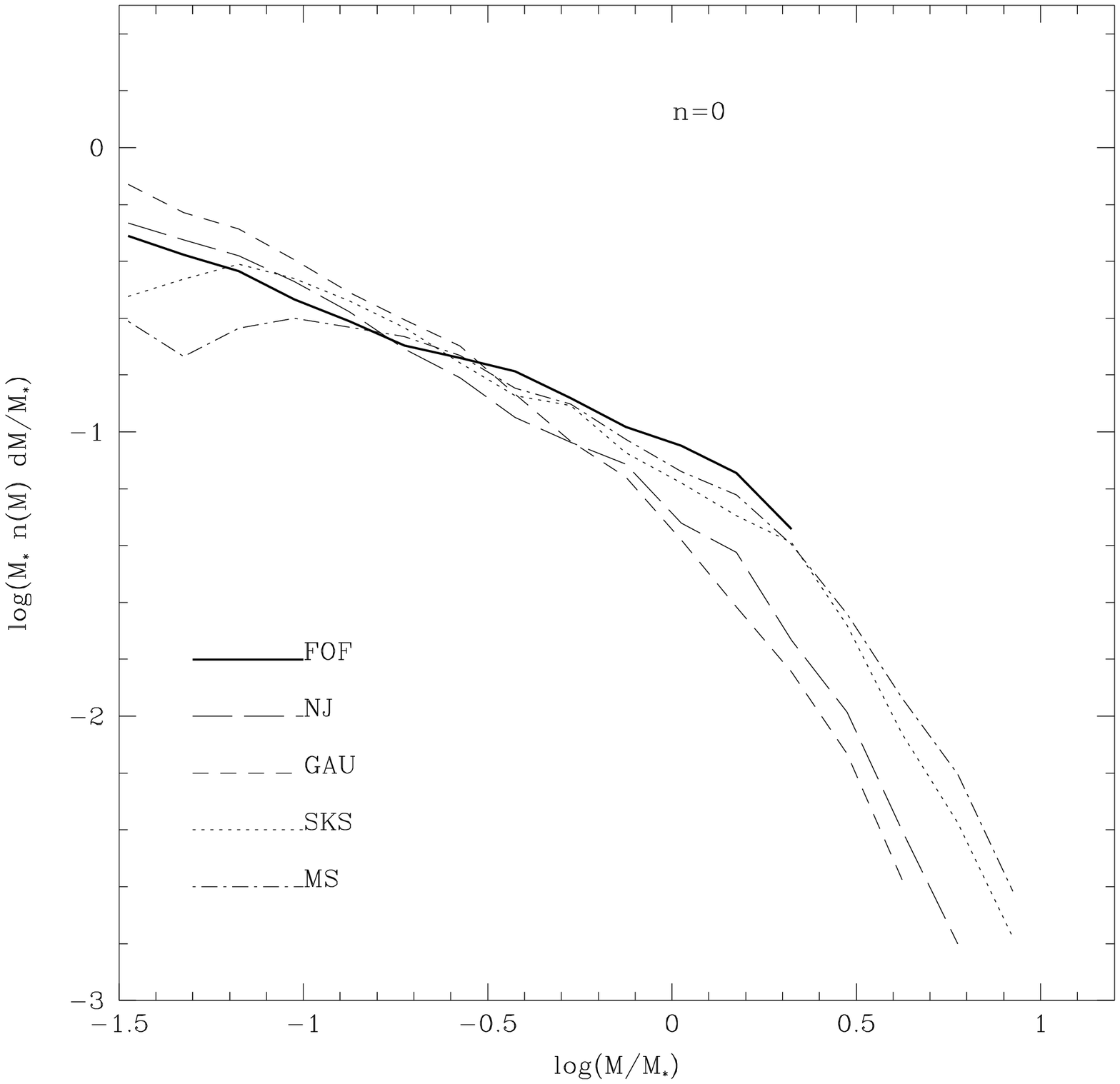,width=7.5cm}
\psfig{figure=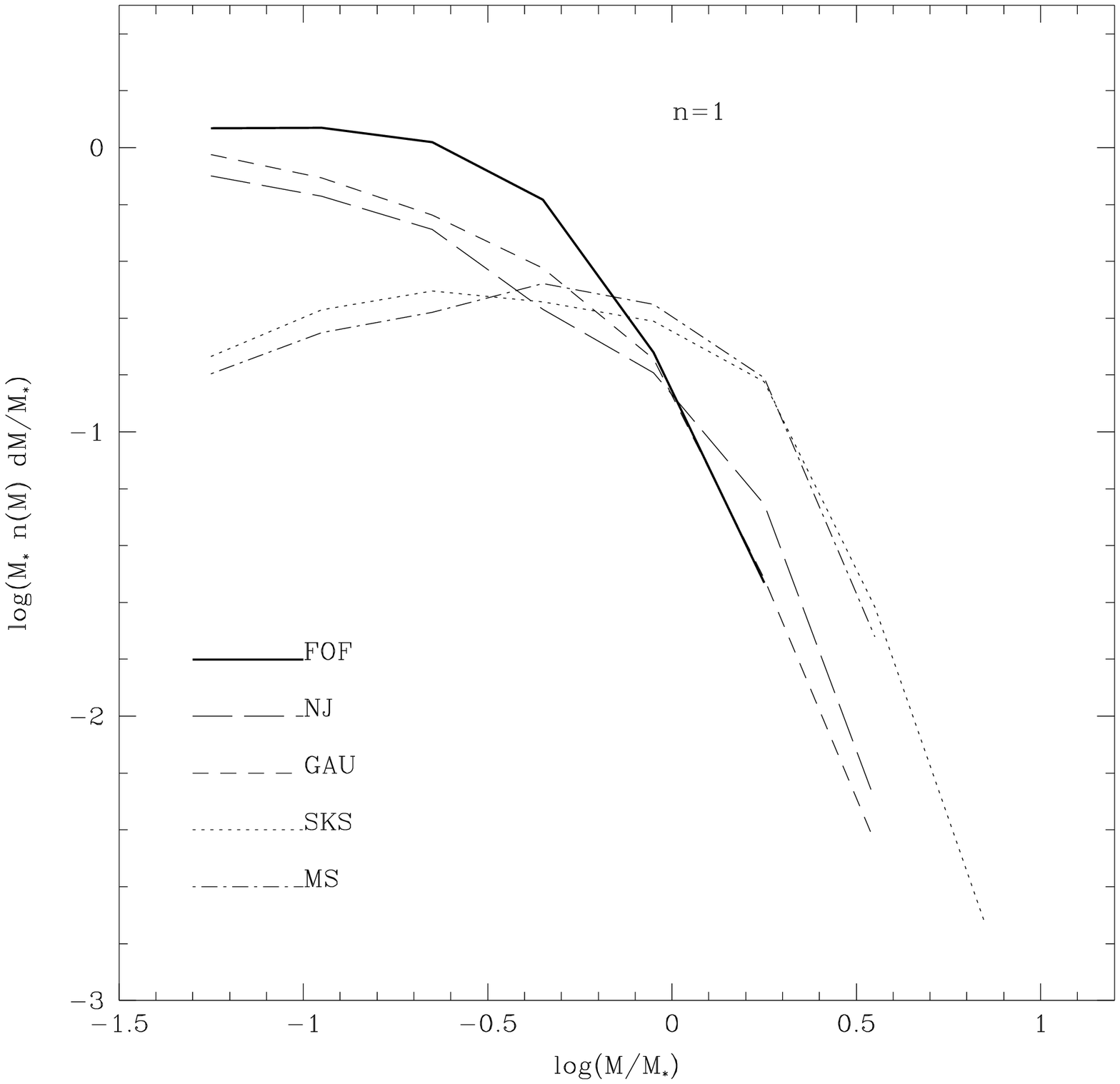,width=7.5cm}
}
\caption{FOF MFs compared to those of Fig. 10.}
\end{figure}

\end{document}